\documentclass{article}

\usepackage{arxiv}

\usepackage[utf8]{inputenc} 
\usepackage[T1]{fontenc}    
\usepackage{hyperref}       
\usepackage{url}            
\usepackage{booktabs}       
\usepackage{amsfonts}       
\usepackage{nicefrac}       
\usepackage{microtype}      
\usepackage{lipsum}
\usepackage{graphicx}
\usepackage{subfigure}
\graphicspath{ {./images/} }
\usepackage{amsmath}
\usepackage{multicol}
\usepackage{multirow}

\title{Novel quantum circuit for image compression utilizing modified Toffoli gate and quantized transformed coefficient alongside a novel reset gate}

\author{
Ershadul Haque, \textit{Member, IEEE} and Manoranjan Paul,  \textit{Senior Member,IEEE} \\
  School of Computing, Mathematics and Engineering\\
  Charles Sturt University\\
  Bathurst, NSW 2795 \\ \textit{Corresponding author email}: 
  \texttt{mhaque@csu.edu.au} \\
}
\begin{document}
\maketitle
\maketitle
\begin{abstract}
Quantum image computing has emerged as a groundbreaking field, revolutionizing how we store and process data at speeds incomparable to classical methods. Nevertheless, as image sizes expand, so does the complexity of qubit connections, posing significant challenges in the efficient representation and compression of quantum images. In response, we introduce a modified Toffoli gate state connection (MTGSC) using a quantized transform coefficient preparation process. This innovative strategy streamlines circuit complexity by modifying state connection from the state connection information. In our operational control gates, only the input '1' impacts the output, allowing us to modify the state connection and dramatically enhance the efficiency of the proposed circuit. As a result, the proposed approach significantly reduces the number of gates required for both image compression and representation.
Our findings reveal that it requires an impressive 44.21\% fewer gates than existing techniques, such as the Direct Cosine Transform Efficient Flexible Representation of Quantum Images (DCTEFRQI), all while maintaining a consistent peak signal-to-noise ratio (PSNR). For an image block size of \(2^{S_x} \times 2^{S_y}\) with \(q\) gray levels, the complexity of our approach can be succinctly expressed as:
\[ O[3q + \log_2{S_x} + \log_2{S_y} + q \cdot 2^{(\log_2{S_x} + \log_2{S_y})}] \]

Here, \(S_x\) and \(S_y\) represent the X and Y positional control gates, while \(q\) indicates the non-zero transform coefficients. Moreover, experimental evaluations strongly demonstrate that it excels in both compressing and representing quantum images compared to the DCTEFRQI approach, particularly excelling in the essential metrics of gate requirements and PSNR performance. Embrace the future of quantum imaging with our innovative solution—where efficiency meets excellence.
\end{abstract}

\section{Introduction}
\label{sec:intro}
Quantum computing is a new branch that combines quantum mechanics, computer science, and mathematics to address the limitations of Moore's law in increasing computing power ~\cite{b6}. It utilizes quantum mechanical properties such as superposition, entanglement, and parallelism to significantly enhance computation speed, storage capacity, and processing complexity for encoding image data~\cite{b6, b5,b1, b2, b3,b4}. In a quantum image, classical bits are replaced by qubits in a pixel array ~\cite{b10}. When image data is represented in a multi-particle photonic system, it is referred to as a quantum image ~\cite{b2}. In the photonic circuit, a qubit serves as the fundamental unit that propagates from left to right, with the image data being placed and connected through each qubit. It can generate every possible processing outcome in detail in terms of hardware processing. Moreover, it can handle an enormous amount of information due to its exponential capacity. Besides, it can automatically generate mathematical ideas or transform data patterns ~\cite{rb7,delp1979image,aaronson2008limits, limits}. 

Nowadays, there is a growing interest in the research community in the field of quantum image compression, representation, and processing due to the faster computation capabilities offered by quantum processors \cite{b5,aaronson2008limits,limits,kalonia2021review}. According to quantum theories in \cite{morvan2023phase, cui2023achieving}, it is much faster than classical computation due to the properties of the quantum mechanics in the Hilbert space. Quantum image processing plays a crucial role in many applications but requires a large amount of memory and hardware resources to handle the vast amounts of data involved. Unlike classical approaches, the representation and compression of quantum images rely on photonic circuit architectures and quantum operational gates. Therefore, to make quantum processors more practical, it is essential to develop highly efficient quantum image compression and representation techniques.

In classical image computing, images are represented as matrix numbers that indicate the color or intensity of each pixel in a Cartesian coordinate system, along with its corresponding location. Spatial correlation is commonly employed in classical image processing, where the location of the pixel is explicitly considered, as it is closely tied to the pixel intensities. However, the traditional compression scheme is inadequate because it does not take position encoding into account. Moreover, conventional image processing methods necessitate substantial computational resources for image processing and storage, unable to handle the problem of non-deterministic polynomial time due to transistor placement restrictions ~\cite{b5}. 

Feynman's exploration of a new type of computer to enhance the computational power of classical computers has generated significant interest in the research community ~\cite{b6}. Consequently, there is a need for new approaches to image representation and compression to address the challenges posed by quantum images. In quantum image computing, qubits are utilized to explicitly encode both pixel intensities and mean positions of the corresponding locations. The quantum circuit only considers the binary representation of '1' for encoding the pixel intensities along with their positions. However, compression in the quantum domain is more complex than in classical systems due to the inclusion of both pixel and its corresponding position. Therefore, efficient schemes must be developed for quantum image representation and compression to optimize memory usage and system costs. 

Inspired by the classic pixel-wise image representation, the NEQR (novel enhanced quantum representation) approach was introduced. Unlike previous methods that approximated pixel values using angles, it directly incorporates the pixel values into the quantum photonic circuit ~\cite{bb10}. In the NEQR circuit, the '1' and '0' connections represent the control and anti-control gate, which is equivalent to the value of the binary bit. However, a limitation of this approach is the inability to represent medium and high-resolution images efficiently. It is mainly due to the individual mapping of each c-not gate, which increases resource requirements and complicates the system. Additionally, the quality of the reconstructed image using this approach has not been evaluated using the required number of gates and PSNR  performance metrics. To overcome the challenges for medium and high-resolution square image mapping, an improved NEQR (INEQR) approach was proposed in a recent study \cite{su2021improved}. However, it is important to note that the outcome of this approach is probabilistic, making it extremely challenging to recover the original image accurately. The reconstruction process from the decoder is also difficult to assess. 

Moreover, a novel approach called the novel quantum representation of color digital image (NCQI) was proposed to enhance the existing NEQR approach for representing color digital quantum images \cite{b16}. However, NCQI is currently limited to a $4\times4$ image size. The representation of medium and high-resolution images is still a challenge due to the pixel mapping strategy to encode each pixel location in the qubit circuit. Additionally, the quality of the reconstructed image at the decoder end has not been investigated due to its non-deterministic nature. 

\par Most of the existing methods directly connect the pixel values representing qubits to the position values representing qubits, where the measurement is done through a probabilistic approach \cite{b8}. The reconstruction of the original image is nearly impossible due to the probabilities-based mapping. As a result, a large number of gates are needed to complete the circuit connection and represent a real image. After that, Nasr et al. proposed an efficient, flexible representation of the quantum image (EFRQI) approach to address the higher number of required gates in the NEQR approach. It achieves higher efficiency by utilizing an auxiliary qubit that establishes a direct connection between the pixel and state-representing qubits \cite{b20}. It also reduces the required number of gates compared to the NEQR approach. However, a significant number of gates and qubit connections are still required due to the repeated use of the same Toffoli gates for each pixel connection. It is only applicable to small-size images due to the complexity of the mapping procedure. Additionally, the non-deterministic approach used does not allow for a quality measure of the reconstructed image to be carried out.

To address the issues of higher qubits  and computational time requirements of the EFRQI approach, a more advanced approach called DCT-EFRQI was developed ~\cite{b9}. It integrates the DCT preparation approach with EFRQI, which resolves the non-deterministic and image size issues. However, it still requires a higher number of gates as it utilizes the EFRQI circuit diagram. Moreover, To tackle the higher required number of gates issue of the DCT-EFRQI approach, a novel technique called SCMFRQI (state connection modification FRQI) was proposed ~\cite{haque2022novel}. It reduces the number of required gates for encoding the grayscale image by utilizing one reset gate instead of using the same Toffoli gate twice for each pixel connection. The efficiency of the SCMFRQI approach is better than the DCT-EFRQI approach. Block-based quantum image is represented with compression in ~\cite{haque2022block} that represents qubit connection value in a block-wise system. It also required a huge amount of gates that included position encoding due to the use of the EFRQI circuit diagram. 

This work proposes a novel MTGSC approach that involves modifying state connections from state-label qubits. Once the $'0'$ connection is discarded, it operates smoothly since the main element that flips the output state of the operational gate is the $'1'$, not the $'0'$. Additionally, the quantum transformation of the $'0'$ can be designed using a quantum identity gate. $'0'$ can be ignored in the quantum circuits, similar to how NEQR ignores it for pixel value preparation ~\cite{bb10}. This ignorance does not impact the overall performance of the quantum circuit. However, it does reduce the complexity of the circuit and improve efficiency in compression as well as representation. 

We choose the MTGSC because:
\begin{itemize}
    \item It can represent the entanglement of the quantized transfer coefficient value and corresponding state label connection using quantum logic gates.  
    \item Due to the use of the quantization process, it becomes a lossy compression technique as well as representation. 
    \item Moreover, due to the use of the block partitioning approach, it requires six qubits to connect $X-$ and $Y-$ state labels in connection to the transform coefficient representing qubits.
    \item Any size of quantum gray-scale images can be compressed and represented using 15 qubits.
    \item It resolves the quantum image's real-time compression and representation issues for gray-scale channel.  
    \item Required number of gates per pixel and PSNR performance measurement criteria are introduced regarding quantum resources.  
\end{itemize}

Since it modifies connection toward getting more compression opportunity in the circuit connection. Therefore, it is not suitable for the application where the modifying state connection is not sacrificed. 

Moreover, the output measure of the existing GQIR becomes a fatal shortcoming due to the probabilistic-based mapping procedure of the pixel value in the Bloch sphere system. Some researchers believe that when one pixel is restored, another pixel is dispersed. To restore the whole pixel image, it needs to be processed, prepared, and measured many times. Due to the use of the deterministic procedure of the SCMNEQR approach, it minimize the fatal issue of GQIR. 

\par The proposed model has been tested and verified using the benchmark dataset and simulator. The experiments were conducted to evaluate the performance of image compression and representation using a quantized DCT-transform approach. The results demonstrate that it outperforms the SCMEFRQI ~\cite{haque2022novel} and EFRQI ~\cite{b9} approaches regarding the required number of transmitted gates savings. The rest of the article is organized as follows: Section \ref{L_R} presents the literature survey, Section \ref{P_M} outlines the proposed methodology, Section \ref{R_D} provides the results and their description, and Section \ref{CC} summarizes the conclusion of this work.

\section{Related Work}
\label{L_R}
In the field of computer vision technology, images from various sectors such as military, medical, and industrial areas are of great importance. Although significant progress has been made in classical image and video compression methods, as discussed in ~\cite{paul2005real} and ~\cite{paul2018efficient} are not suitable for quantum image representation and compression. However, the emergence of quantum computing has opened up new possibilities for quantum image processing (QIP) ~\cite{b11}. It employs a quantum concept framework that utilizes quantum logic gates to perform image operations and address storage issues associated with classical images. By leveraging the power of quantum state superposition and entanglement theories, QIP greatly enhances image processing capabilities. Notably, within the realm of QIP, quantum computing technology ~\cite{yan2017quantum}, image watermarking ~\cite{song2014dynamic, song2013dynamic, iliyasu2012watermarking, hu2019quantum, kong2010color}, image compression ~\cite{b13, pm}, image storage and retrieval ~\cite{li2014multi}, image encryption, quantum security~\cite{peelam2024explorative}, among others, have shown significant potential ~\cite{wang2022efficient, wang2014novel}. In quantum encryption, post-quantum security becomes another rising field~\cite{khan2024chaotic}. The artificial intelligence-based quantum Synergies approach is another promising field of quantum computing in software-defined consumer applications~\cite{awan2024artificial,wu2022state}. 

First, to address the issue of representing images on a quantum processor, a qubit lattice approach is used to represent the four different random pixels of an image using a single qubit ~\cite{b10}. It is the first approach to demonstrate the storage and retrieval of quantum images in a multi-particle quantum system. However, due to the use of a single qubit, it is difficult to represent both the image pixels and their positions. The real-ket approach is another approach to represent quantum images, which utilizes the ket function of the qubit ~\cite{b12}. However, it has limitations when it comes to representing larger, real-size images in the qubit system due to the limited number of pixel values such as four. In the rapidly evolving field of computer vision technology, the significance of images from diverse sectors—including military, medical, and industrial—cannot be overstated. While remarkable advancements have been achieved in classical image and video compression techniques, as highlighted in ~\cite{paul2005real} and ~\cite{paul2018efficient}, these conventional methods fall short when it comes to quantum image representation and compression. The advent of quantum computing has ushered in groundbreaking opportunities for quantum image processing (QIP)~\cite{b11}. By harnessing a quantum conceptual framework and employing quantum logic gates, QIP performs sophisticated image operations and effectively tackles the storage challenges that classical methods face. Through the utilization of quantum state superposition and entanglement, QIP dramatically elevates image processing capabilities to new heights.

Within this dynamic realm, QIP has demonstrated substantial potential across various applications, including quantum computing technology ~\cite{yan2017quantum}, image watermarking ~\cite{song2014dynamic, song2013dynamic,iliyasu2012watermarking, hu2019quantum, kong2010color}, image compression ~\cite{b13, pm}, image storage and retrieval ~\cite{li2014multi}, and image encryption ~\cite{wang2022efficient, wang2014novel}. These advancements reveal the true power of quantum-based methodologies.

To effectively tackle the challenge of representing images on quantum processors, innovative approaches are being explored. The qubit lattice approach, for instance, allows the representation of four distinct random pixels of an image using a single qubit ~\cite{b10}. This groundbreaking method is the first to showcase the storage and retrieval of quantum images in a multi-particle quantum system. Nonetheless, the limitation of using a single qubit complicates the representation of both image pixels and their spatial positioning.

Another promising approach, known as the real-ket method, leverages the ket function of qubits to represent quantum images ~\cite{b12}. However, this method too encounters challenges when it comes to representing larger, real-world images due to the confined number of pixel values it can handle, specifically four. It is crucial to continue exploring these avenues to fully unlock the transformative potential of quantum image processing.

\begin{figure}[htbp]
\centerline{\includegraphics[width=0.30\linewidth,height=0.30\linewidth]{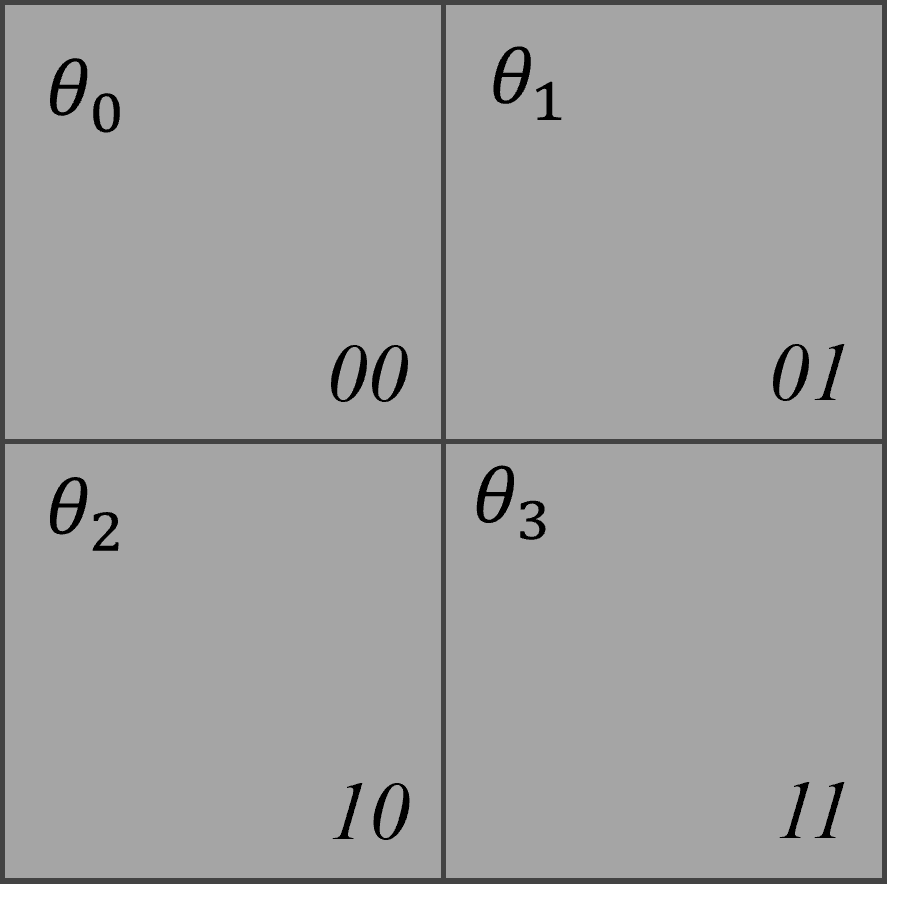}}
\caption{A $2\times2$ FRQI quantum image}
\label{FRQI_A}
\end{figure}

Shor and Grover proposed a faster approach than classical computation for factorial calculation and database search ~\cite{b5, b6}. In 2011, an FRQI (flexible representation of the quantum image) approach was Shor and Grover proposed a faster method for factorial calculation and database searching compared to classical computation ~\cite{b5, b6}. In 2011, a Flexible Representation of the Quantum Image (FRQI) approach was introduced, which takes into account the probability of pixel values and normalizes the state for mapping ~\cite{b13}. However, this approach has a significant limitation: it relies on probabilistic outcomes. Additionally, a quality measurement of the reconstructed image is necessary. Figure~\ref{FRQI_A} illustrates the quantum image for the FRQI approach, showcasing its state for a \(2\times2\) image which considers the probability of pixel values and normalizes the state for mapping~\cite{b13}. However, it has a significant limitation: it relies on probabilistic outcomes—furthermore, the quality measurement of the reconstructed image needed to be conducted. Figure~\ref{FRQI_A} depicts the quantum image for the FRQI approach, including its state for a $2\times2$ image. 

The equation below represents the mathematical expression of the FRQI image, $|I_{FRQI}\rangle$.\\

\begin{math}
|I_{FRQI}\rangle=\frac{1}{2} [\left(cos\theta_{0}|0\rangle +sin\theta_{0}|1\rangle \right) \otimes |00\rangle 
+\left(cos\theta_{1}|0\rangle+sin\theta_{1}|1\rangle \right)\otimes|01\rangle +  \left(cos\theta_{2}|0\rangle+sin\theta_{2}|1\rangle \right)\otimes|10\rangle+ \left(cos\theta_{3}|0\rangle+sin\theta_{3}|1\rangle \right)\otimes|11\rangle ]\nonumber
\end{math}

\begin{figure}[t!]
\centerline{\includegraphics[width=0.8\linewidth,height=4cm]{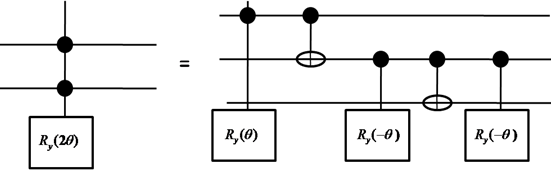}}
\caption{An FRQI circuit for representing $2\times2$ image}
\label{FRQI_Circuit}
\end{figure}

Figure~\ref{FRQI_Circuit} shows FRQI circuit of the image $|I_{FRQI}\rangle$.  

The formula for the standard rotation metric, $R_y(2\theta)$, is expressed as follows, 
$\begin{pmatrix}
\centering
  cos\theta_i & -sin\theta_i \\ 
  sin\theta_i & cos\theta_i
\end{pmatrix}$ 

Where $\theta$ is the angle of the corresponding qubits, it represents the image pixel values in the quantum domain using the control rotation matrices. The control rotational matrices were implemented using a c-not gate and standard rotation. It cannot represent pixel-wise gray-scale complex operations since it uses only one qubit. It encodes the color and position of the image using the associated angle and kets of one qubit. It can only represent four numbers of pixel values of the image. Generally, it uses rotating gates to store gray-scale pixel values in the probabilistic-based amplitude mapping in the Bloch sphere. Bloch sphere is the geometry of a vector, whose unitary transformation is a rotating matrix indicating the magnitude of the state purity ~\cite{gamel2016entangled}. 

An entanglement-based image representation approach was proposed using the image entanglement theory to represent image data in the quantum circuit ~\cite{b14}. It removes any additional required information, such as correlation. Additionally, it can convert image data into a normalized state using a reduced number of operations. It maps the pixel value directly into an angle for a binary image. More sufficient details about quantum image knowledge should have been provided. Moreover, it is limited to small sizes of images, which are not suitable for real-life applications.

A NEQR approach was proposed to address the classical gray-scale image representation issue of the FRQI method ~\cite{bb10}. It resolves the FRQI representation issue because it provides a way to represent the pixel-based representation of the gray-scale image. After converting the pixel values into a binary system, only a frequent number of ones are considered to map the pixel value in a quantum system. The pixel and state (position) representing qubits map each pixel and the corresponding position values. For example, an image whose pixel values are $0 (Y=0, X=0), 100 (Y=0, X=1), 200 (Y=1, X=0), 255 (Y=1, X=1)$ and its respective quantum representations are known as the NEQR approach and express as,

\begin{flalign}
\resizebox{0.9\hsize}{!}{$
|I_{NEQR}\rangle=\frac{1}{2}[|0\rangle\otimes|00\rangle 
+|100\rangle\otimes|01\rangle+|200\rangle\otimes|10\rangle+|255\rangle \otimes|11\rangle]\nonumber
$}
\end{flalign}

\begin{figure*}[htbp]
\centerline{\includegraphics[width=\linewidth,height=6cm]{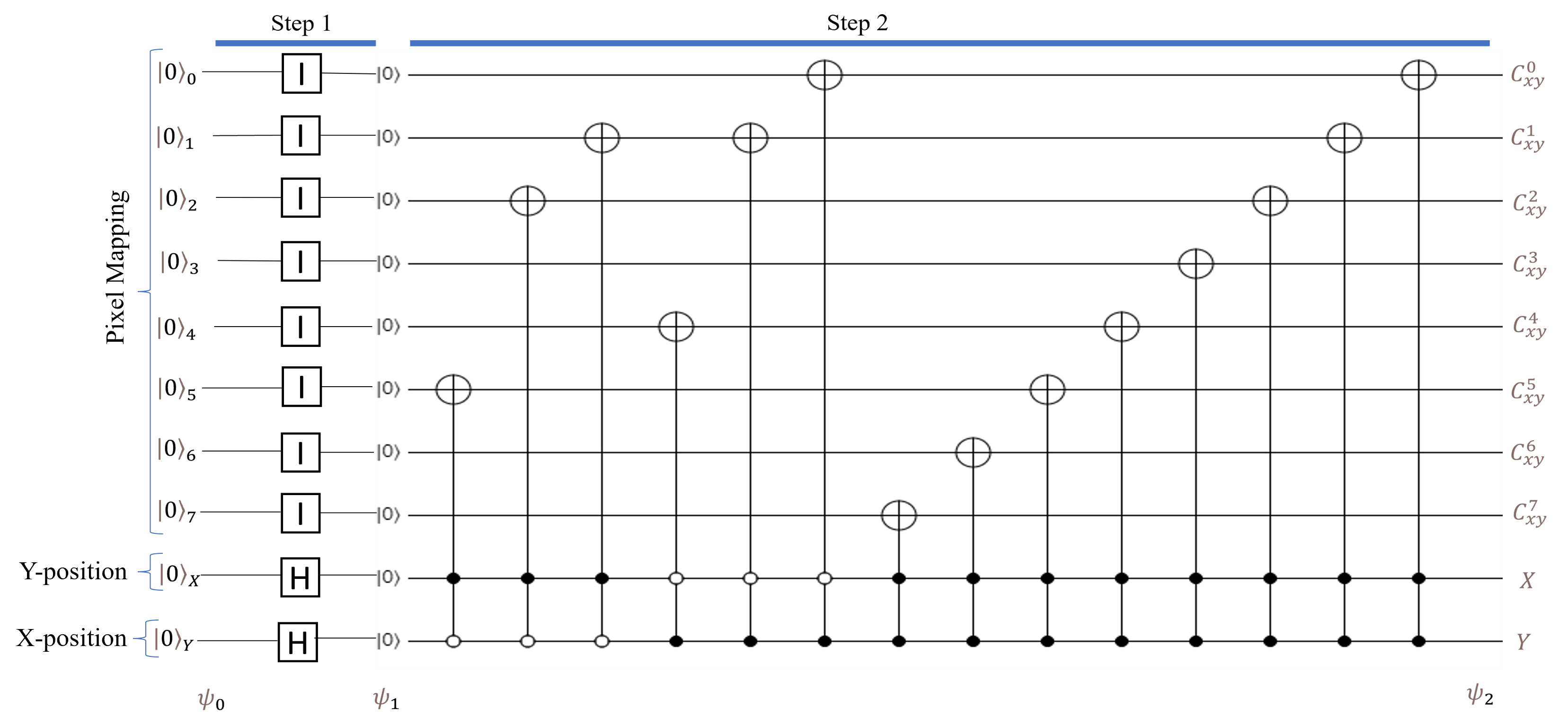}}
\caption{An NEQR circuit diagram for pixel values representation}
\label{NEQR_Circuit}
\end{figure*}

Figure~\ref{NEQR_Circuit} depicts the circuit diagram for the $|I_{NEQR}\rangle$ image, which effectively addresses the FRQI issues by utilizing multiple qubits to encode the gray-scale image pixels. Nonetheless, several limitations exist associated with this approach. Firstly, it is incapable of representing medium or high-resolution images that exceed the dimensions of $4\times4$ due to the increased connectivity required for additional qubits. Furthermore, the outcome measurement approach is probabilistic, which significantly hampers the accurate restoration of the original image. Moreover, images with rectangular shapes cannot be accurately represented.

For real-size images, how and in what way to represent them still needs to be clarified. INEQR (improved NEQR) approach was proposed in ~\cite{b9} to resolve the rectangular shape images issue of the NEQR approach; it still needs to clarify how the color and large-size images are represented. Also, it is only able to represent images of very tiny sizes. Moreover, the mapping procedure is probabilistic. As a result, the measured reconstructed image quality needs to be performed due to the non-deterministic outcome. 

In 2018, Li et al. introduced an Optimized Quantum Representation for Color Images (OCQR) approach, which focuses on the representation of color image pixels ~\cite{liu2018optimized}. This method requires more qubits than the NEQR approach but utilizes fewer qubits compared to the NCQI approach, due to the incorporation of index values. However, it is limited to image sizes of $2 \times 2$. In the same year, Nanrun et al.(2018) proposed a Quantum Representation Model for Multiple Images (QRMMI) ~\cite{zhou2018multi}. This model aims to represent multiple images simultaneously, thereby conserving hardware space for image representation. However, it relies on the NEQR approach for the state-label circuit connection, which necessitates a large number of qubit connections, leading to increased circuit complexity.

\begin{figure*}[htbp]
\centerline{\includegraphics[width=\linewidth,height=8cm]{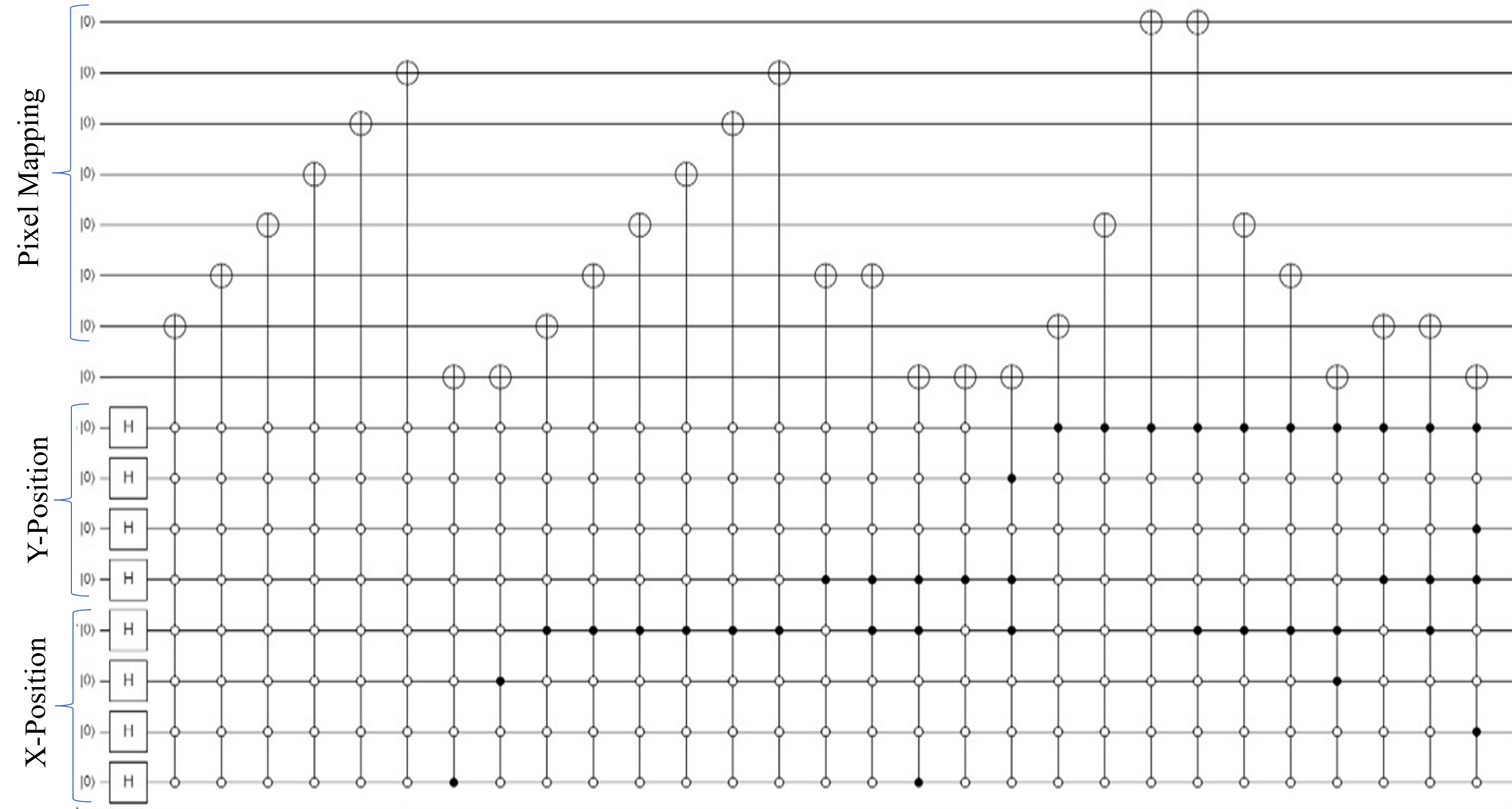}}
\caption{A GQIR circuit for representing pixel values}
\label{GQIR_Circuit}
\end{figure*}

Following this, a General Quantum Image Representation (GQIR) approach was presented in \cite{b18}. This approach uses a logarithmic scale to represent images of both square and rectangular sizes. For an image size of $512 \times 512$, it requires nine qubits to encode the $Y$-positions and an additional nine qubits for the $X$-positions. Figure~\ref{GQIR_Circuit} illustrates a quantum circuit example for a $16 \times 16$ image of a deer, which is generated after applying the DCT preparation approach and utilizing 70 quantization factors with the Quirk simulation tool~\cite{bb17}. The transform coefficient values for $16\times 16$ size deer image  are, $126(X = 1, Y = 1)$, $1(X = 1, Y = 0)$, $1(X = 4, Y = 0)$, $126(X = 8, Y = 0)$, $4(X = 0, Y = 1)$, $1(X = 8, Y = 1)$, $1(X = 9, Y = 1)$, $1(X = 8, Y = 1)$, $1(X = 8, Y = 5)$, $138(X = 0, Y = 8)$, $140(X = 8, Y = 8)$, $1(X = 12, Y = 8)$, $2(X = 0, Y = 9)$, $2(X = 8, Y = 9)$ and $1(X = 2, Y = 11)$. 

The use of zero for connectivity results in the generation of redundant qubits connection. Additionally, representing medium to high-resolution images using a GQIR circuit is challenging due to the probabilistic mapping involved. Consequently, reconstructing the original image after encoding in the quantum domain is nearly impossible because of its non-deterministic outcomes. Furthermore, Jiang et al. proposed a compression method that combines the GQIR and DCT preparation approaches based on the Jpeg (Joint Photographic Experts Group) standard. Another study introduced a quantum-based equivalence pixel image derived from a bit pixel image. The use of zero for connectivity leads to the generation of redundant qubits connection. Additionally, representing medium or higher-resolution images using a GQIR circuit is challenging because of the probabilistic mapping. It is nearly impossible to reconstruct the original image after encoding in the quantum domain due to non-deterministic outcomes. Moreover, Jiang et al. proposed a compression approach that utilizes the GQIR and DCT preparation approaches based on JPEG (joint photography expert group) image~\cite{pm}. Another study proposed a quantum-based equivalence pixel image from a bit pixel image~\cite{b15}.

\par
\begin{figure*}[htbp]
\centerline{\includegraphics[width=\linewidth,height=8cm]{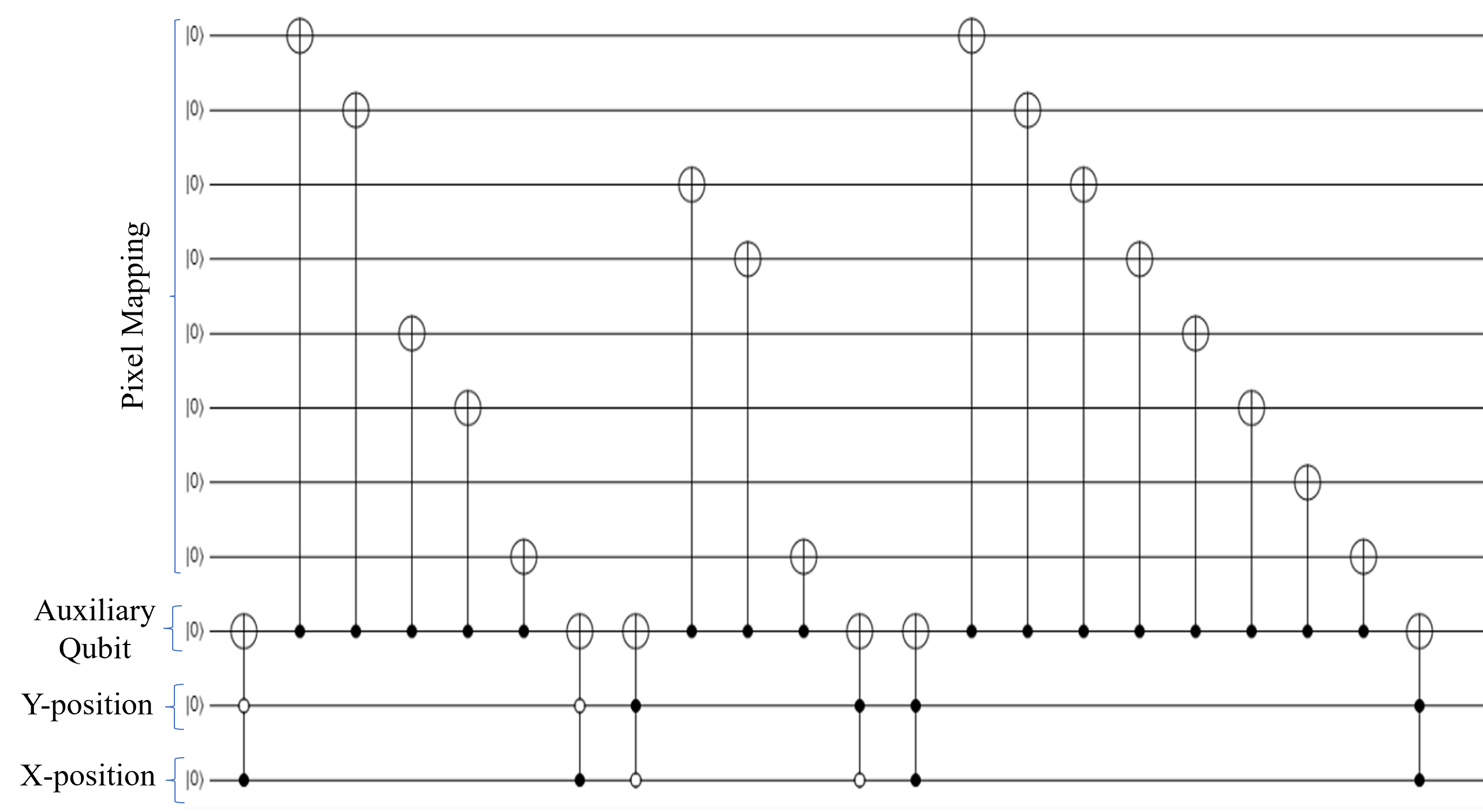}}
\caption{An EFRQI circuit for pixel values representation}
\label{EFRQI_circuit}
\end{figure*}

 An efficient flexible representation of the quantum image (EFRQI) was proposed to decrease the state preparation complexity of GQIR and NEQR approaches~\cite{b20}. Figure~\ref{EFRQI_circuit} shows the EFRQI circuit of the gray-scale image whose pixel values are $205(X=1, Y=0)$,$49(X=0, Y=1)$, and $255(X=1, Y=1)$. Each pixel value connection uses two Toffoli gates, one for initiating pixel connection and another for closing. The benefit of this approach is that it does not depend on the number one that occurs in each pixel for pixel position connection. As a result, it saves the position values more than the GQIR approach. Due to the use of the same Toffoli gate twice for each pixel connection, it generates a higher amount of required qubits connection. To address the higher required quantum resources issue of the EFRQI approach, Haque et al. (2022)~\cite{b9} proposed a DCTEFRQI approach that integrates the DCT approach with the EFRQI circuit. It outperforms the EFRQI approach. Plenty of gates are still required due to the use of Toffoli gates. 

\par
\begin{figure*}[htbp]
\centerline{\includegraphics[width=\linewidth,height=0.45\linewidth]{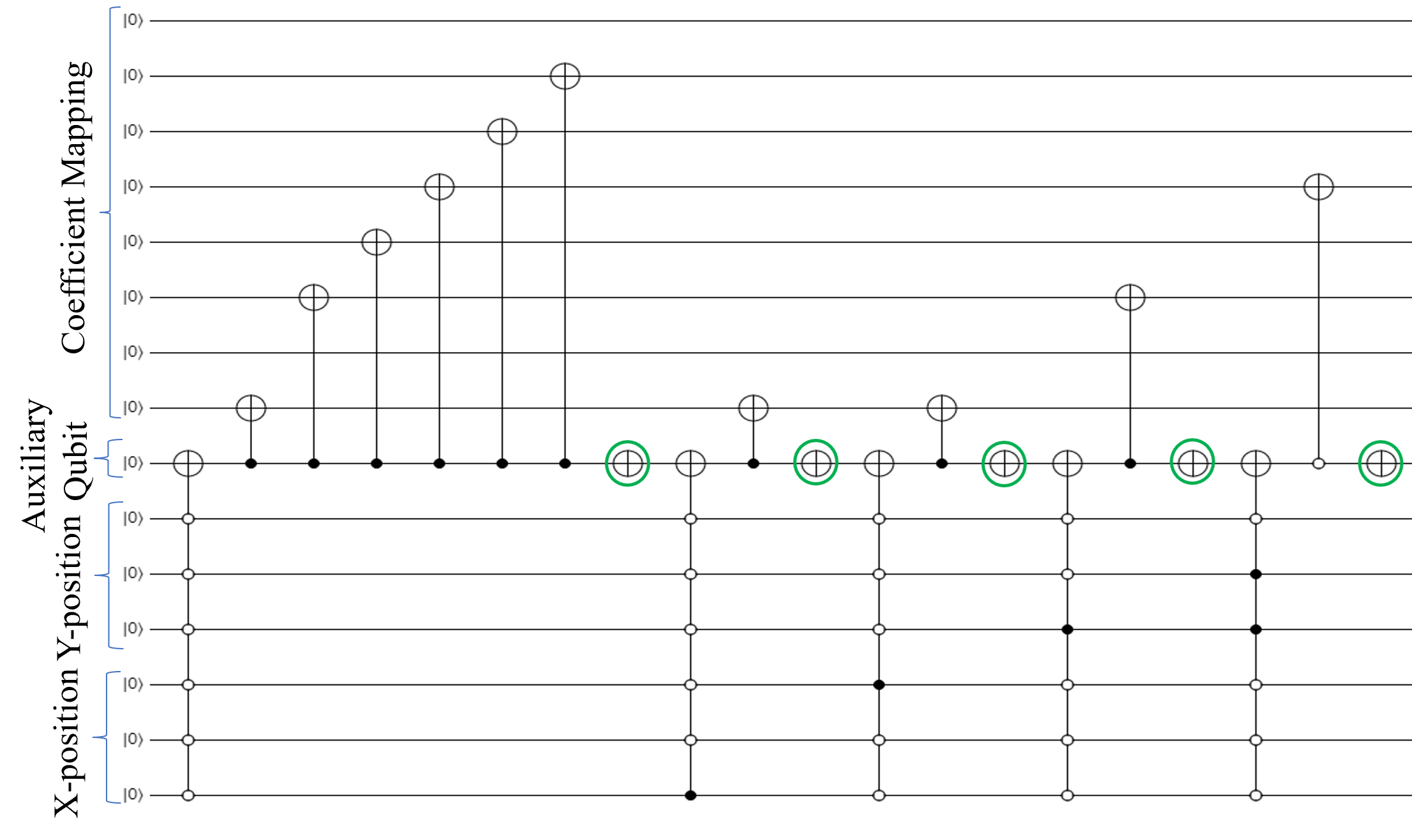}}
\caption{A SCMFRQI circuit for transformed coefficient value representation.}
\label{fig_proposed_Chematic_diagra}
\end{figure*}

 In~\cite{bb17}, a novel reset gate called the SCMFRQI approach is introduced as a replacement for the closing Toffoli gate connection in the EFRQI circuit. Figure~\ref{fig_proposed_Chematic_diagra} illustrates the circuit of this approach, which represents transform coefficient values of $125(X=0, Y=0), 1(X=1, Y=0), 1(X=4, Y=0), 4(X=0, Y=1),$ and $16(Y=3, X=0)$ ~\cite{haque2022novel}. Instead of using the Toffoli gate twice, it utilizes a single reset gate for each coefficient connection (shown in the green circle) ~\cite{bb17}. Consequently, the required number of gates is reduced compared to the DCTEFRQI approach.

\section{Proposed approach}
\label{P_M}
In this section, we provide a detailed description of the proposed approach. As the use of quantum computers continues to grow, the need to represent and store large amounts of data has become increasingly important. For instance, individuals managing a webpage or online catalog often utilize dozens or even hundreds of images. Today, there are numerous available image compression techniques, which can be classified into two main categories: lossy and lossless compression. Jpeg is a widely-used image compression scheme that employs the discrete cosine transform (DCT) as its transformation method. In JPEG, DCT serves as a pre-processing step.

\begin{figure*}[!t]
\centerline{\includegraphics[width=\linewidth,height=8cm]{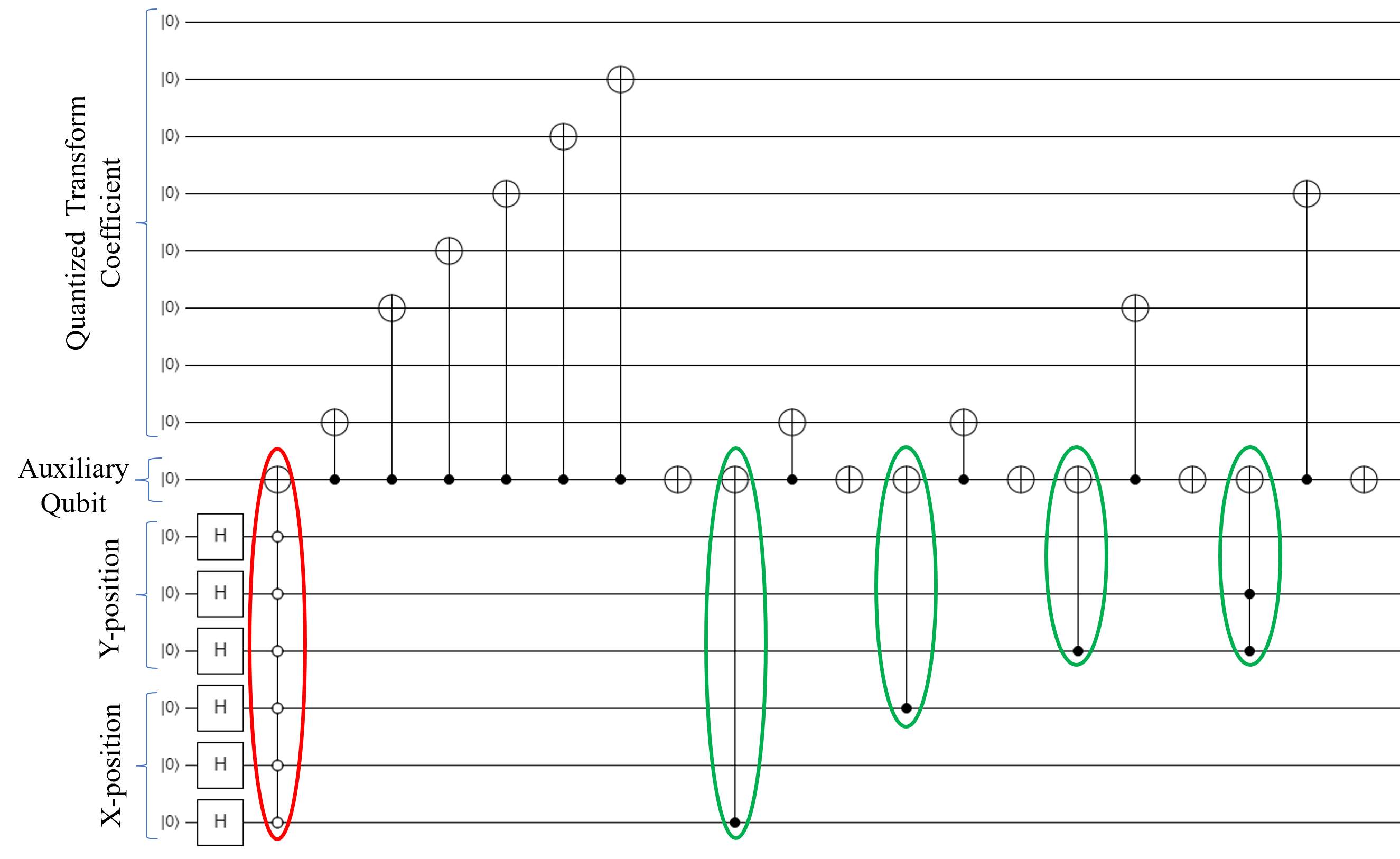}}
\caption{Proposed MTGSC circuit for quantized transform coefficient and position values mapping for compression. It includes an initial connection (marked in red) and a state modification zone (marked in green).}
\label{fig_proposed_SCMNEQR_diagram}
\end{figure*}

Figure~\ref{fig_proposed_SCMNEQR_diagram} illustrates the circuit of the proposed MTGSC scheme for mapping coefficient values of $125(X=0, Y=0)$, $1(X=1, Y=0)$, $1(X=4, Y=0)$, $4(X=0, Y=1)$, and $16(Y=3, X=0)$ for compression. Rather than using both $'0'$ and $'1'$ for positional connections, this approach utilizes $'1'$ exclusively for each coefficient connection (as shown in the green circle). 

\begin{figure*}
    \centering
    \subfigure[]
    {
        \includegraphics[width=\textwidth, height=10cm ]{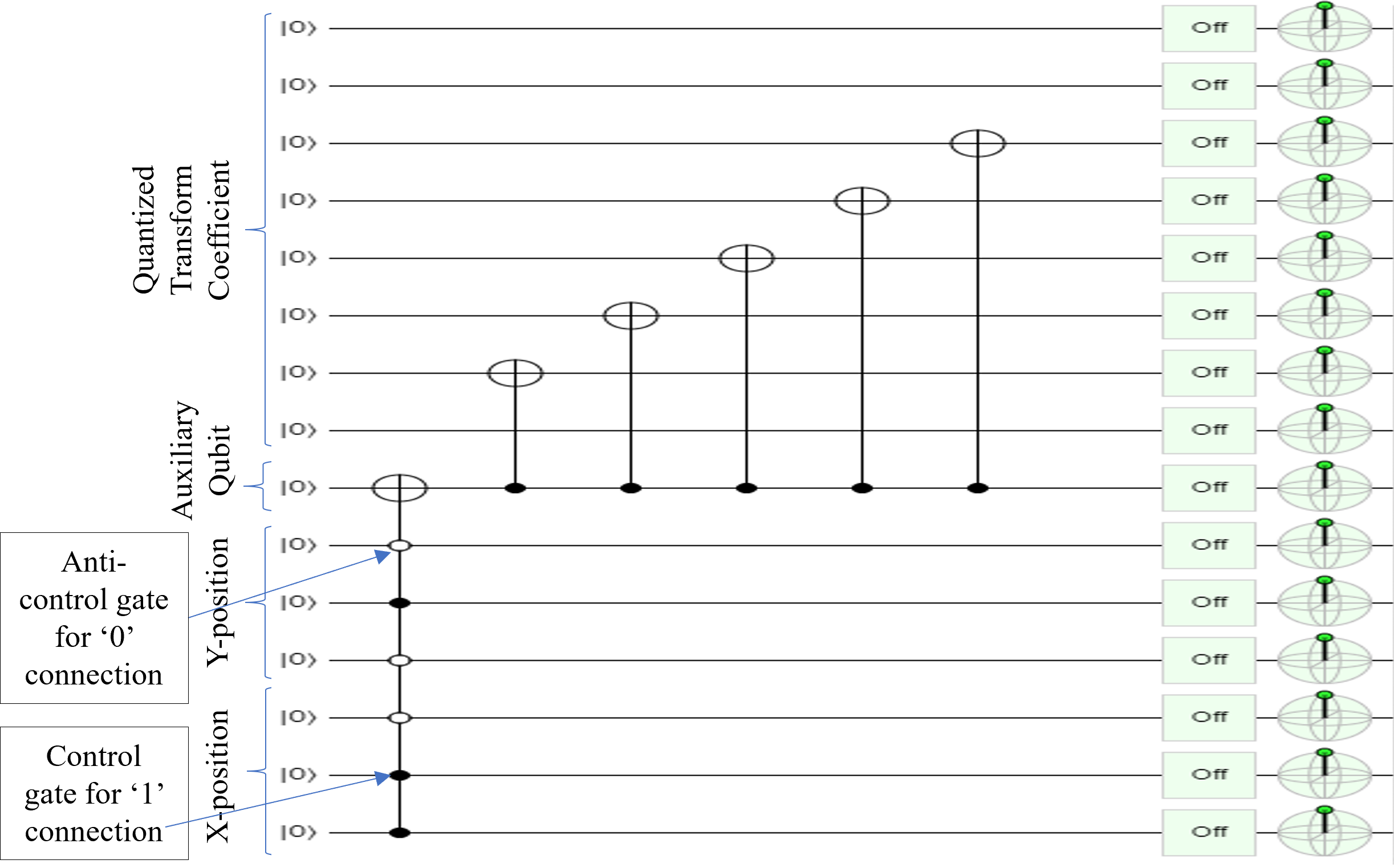}
        \label{Ex_with_zero}
    }
    \subfigure[]
    {
        \includegraphics[width=\textwidth, height=10cm ]{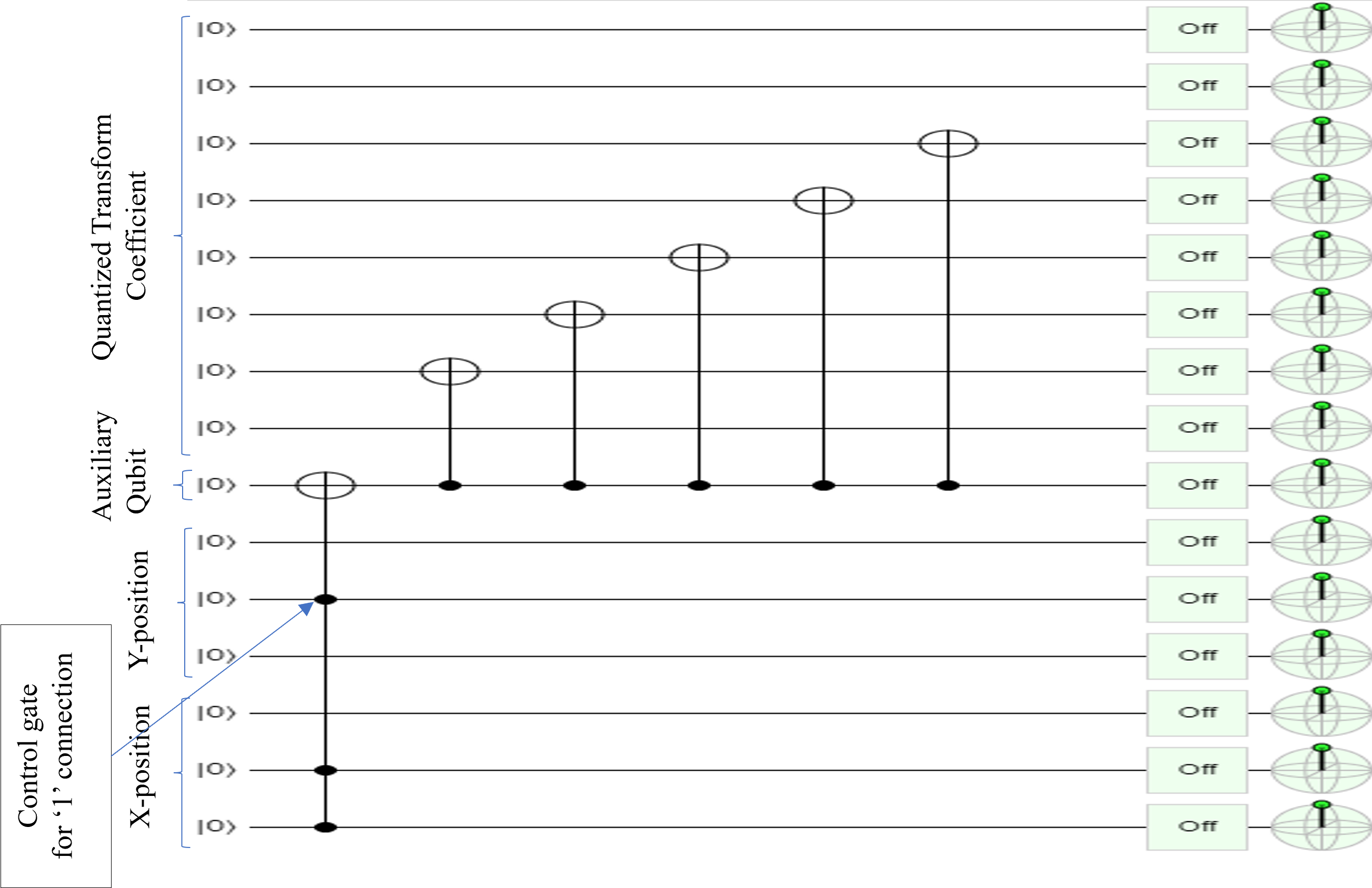}
        \label{Ex_without_zero}
    }
    \caption{Demonstration of state connection modification effect to prepare quantized transform coefficient circuit of the proposed MTGSC approach.}
    \label{Example_1}
\end{figure*}

For example, Figure~\ref{Ex_with_zero} demonstrates the preparation of the quantized transform coefficient with the value $62(X=3,Y=2)$. The results indicate no change in the amplitude of the probability distribution. Furthermore, Figure~\ref{Ex_without_zero} illustrates the effect of state connection modification from the positional preparation. This measurement result suggests that modifying state connections does not affect the outcome, as they function similarly to an identity gate. This analysis supports the feasibility of modifying state connections from the positional preparation, consistent with the approach used for NEQR pixel preparation in ~\cite{bb10}.

Quirk simulation tool has been used to map and visualize quantum circuits and their concept~\cite{bb17}. 

A $2^a \times 2^b$ image size is considered for representation and compression purposes. 
In the proposed approach, the steps involved are,\\ 
Step 1: DCT and quantize. \\
DCT divides the image into $8\times 8$ blocks of pixels and then applies to each block from left to right to get the $F(x,y)$ frequency spectrum. 

\begin{align}
F(x,y)=C(x)C(y) \sum_{i=0}^{7}\sum_{j=0}^{7}\,\bigl\{ P(i,j) cos \left[ \frac{(i+0.5)\pi}{8}i\right] \\ cos \left[ \frac{(j+0.5)\pi}{8}j\right]\bigl\}\nonumber
\end{align}

where $i=0,1,...7$, $j=0,1,...7$ and 

\begin{equation}
  C(x) =
    \begin{cases}
      \frac{1}{2\sqrt{2}} & \text{if $x=0$}\\
      \frac{1}{2} & \text{if $x\not =0$}\\
     \end{cases}       
\end{equation}

The higher the values of \( x \) and \( y \), the greater the frequency spectrum \( F(x, y) \). Additionally, \( F(0, 0) \) represents the direct current (DC) coefficient value. The human eye is not very adept at distinguishing between high-frequency differences, which allows for significant high-frequency reduction. This reduction is achieved through a process called quantization. 

In quantization, a scalar quantization factor is used to divide the transform coefficients, which are then rounded to the nearest integer. This rounding operation is a lossy step in the entire process, resulting in many higher-frequency components being rounded to zero. Meanwhile, the remaining frequency components can be either positive or negative, requiring fewer qubits connection for transmission. 

The quantization factor regulates the compression ratio: a higher quantization factor leads to greater compression. Following this, the quantized coefficient values of the Discrete Cosine Transform (DCT) are calculated. 
\begin{equation}
    F_Q(x,y)=round\Big(\frac{F(x,y)}{Q} \Big)
\end{equation}

Moreover, sign bits are computed with: 
\begin{equation}
 F_Q{_{sign}}=sign(F_Q(x,y))
\end{equation}

$Q$ is considered a scaler quantization value of 2, 4, 8, 16, 32, 36, 60, 70, 90 and 120. 

Step 2: Prepare a quantized DCT coefficient. The proposed MTGSC approach requires $(q+n+1)$ qubits, where $q$ is the number of required qubits to represent the coefficient values and $n=\log_2(S)$. Here, $S$ is the total number of blocks consisting of $XY-$ positional blocks of the images. The initial state can be explained using the equation below ~\cite{b20}.

 \begin{equation}
   |\Psi_0\rangle={\vert0\rangle}^{\otimes(q+n+1)}
 \end{equation}
 
Then, $(q+1)$ identity gates and $n$ hadamard gates are used. The entire quantum preparation procedure can be expressed as follows, 

\begin{equation}
U=I^{\otimes{q+1}}\otimes H^{\otimes{n}}
\label{eq6}
\end{equation}

$U$ transforms $\Psi_0$ from the initial state to the intermediate state $\psi_1$.
\begin{equation}
\Psi_1=U(|\Psi_0\rangle)=(I|0\rangle)^{\otimes{q+1}}\otimes (H|0\rangle)^{\otimes{n}}
\end{equation}
The final preparation step ($\Psi_2$) uses the $U_2$ quantum operator. The $\Psi_2$ operator is given as, 
\begin{flalign}
\Psi_2&=U_2(|\Psi_1\rangle) \\  
&=\frac{1}{2^n} \sum_{i=1}^{n-1}\sum_{j=1}^{n-1}\,\left(|C_{YX}\rangle (|Y_{o}X_{o}\rangle \right)
+\frac{1}{2^n}|C_{Y_zX_z}\rangle|Y_{z}X_{z}\rangle) \nonumber
\end{flalign}
Where $|C_{YX}\rangle$ is the corresponding coefficient value of the $Y_{o}X_{o}$ position, which considers the values of ones only. On the other hand, $|C_{Y_zX_z}\rangle$ represents the coefficient value of the zero position, and $Y_{z}X_{z}$ represents its remaining positional value. The quantum transform operator $U_2$ is given as,
\begin{equation}
U_2=\prod_{X=0,....,2^n-1}\prod_{Y=0,....,2^n-1}\, \left(U_{Y_{z}X_{z}}+U_{Y_{o}X_{o}}\right)
\end{equation}
The quantum sub-operator $U_{YX}$ is expressed as,  
\begin{flalign}
U_{YX}&= \left(I\otimes \sum_{ij\neq YX} {|ji\rangle {\langle ji|}} \right) +\sigma_{YX} \otimes|YX\rangle {\langle YX|}
\end{flalign}
where $|YX\rangle {\langle YX|}=|Y_{z}X_{z}\rangle {\langle Y_{0}X_{0}|}+|Y_{0}X_{0}\rangle {\langle Y_{0}X_{0}|}$. 
The $\sigma_{YX}$ is given below, 
\begin{equation}
    \sigma_{YX} =\otimes^{q-1}_{i=0}{\sigma^i_{YX}}
\end{equation}
The function of $\sigma^i_{YX}$ is setting the value of $i_{th}$ qubit of $(YX)'$s quantized DCT coefficient. \\
Step 3: Store the quantized coefficient after performing $8\times8$ block and compute the required number of gates.\\
Step 4: Compute the required number of gates and PSNR performance metric. \\
The required number of qubits connection of Toffoli gate $(B_{T})$ connection is,       
 \begin{equation}
B_{T}= (log_2(S_X)+log_2(S_Y)+C_T)\otimes {N_{tcn}}
 \end{equation}
The required number of reset gates ($B_{rg}$) is given as, 
 \begin{equation}
          B_{rg}= R_N\otimes{N_{tcn}}
 \end{equation}
Where $R_N$ is the required number of connections. The total number of required quantum gates is calculated, with zero ($B_{z}$) as follows, 
 \begin{equation}
          B_{s_0}= B_{T}+B_{rg}-B_{z}
 \end{equation}
 In the proposed circuit, zero (0 volts) and one (5 volts) are considered as binary bits. Modifying state connections makes the proposed circuit more efficient because it requires less operational gates. A considerable number of operational gates is required if '0' is considered to complete the entire image circuit. Therefore, in the proposed method, modify state connection saves a significant number of operational gates and has no effect since it is implemented by a quantum identity gate, which can be ignored.
 The calculation for the total number of required gates ${BR|_{trg}}$ is as follows,  
 \begin{equation}  {BR|_{MB}}=\left(q_{o}+S_{bit}+B_{s_0}+A_{bit}+B_{BPE}\right)/(1000\times 1000)
 \end{equation}
\begin{equation}
     {BR|_{BPE}}=N_{rb}\times N_{cb} \times (B_{rbc}+B_{rbr})/(1000\times 1000)
 \end{equation}

 \begin{equation}  
 {BR|_{trg}}={BR|_{MB}}/(S_i)
 \end{equation}
 
Where $q_{o}$ represents the total number of ones in the transfer coefficient values, $S_{bit}$ is the sign bit indicating the sign of non-zero coefficient values, and $N_{tcn}$ is the total number of non-zero coefficient. $S_X$ and $S_Y$ denote the $X$ and $Y$ positions of non-zero coefficients. An $A_{bit}$ is the number of required qubits from the auxiliary qubits. $BR|_{BPE}$ stands for the block positional error used to locate blocks beyond the first one. $B_{rbc}$ and $B_{rbr}$ are the number of qubits required to locate the missing row ($N_{rb}$) and column blocks ($N_{cb}$) respectively. $BR|_{MB}$ represents the required number of gates, and $S_i$ refers to the image size. $C_T$ indicates the number of reset gates needed to complete the coefficient representing the circuit diagram.\\
Step 5: Inverse quantization. \\
Step 6: Inverse procedure of the DCT approach is executed as follows:\\
\begin{enumerate}
    \item After compression, from the $8\times8$ block of the image, reform the compressed transfer coefficient. After that, multiply the compressed transform coefficient value by the sign value.
    \item Then, multiply each $8\times8$ block which includes the sign value.
    \begin{equation}
    F(i,j)=F_Q{_{sign}}\times Q 
    \end{equation}
\end{enumerate}

To get the reconstructed image $F'(i,j)$, apply inverse DCT to each $8\times8$,

\begin{align}
F'(i,j)=C(x)C(y) \sum_{i=0}^{7}\sum_{j=0}^{7}\,\bigl\{ F(i,j) cos \left[ \frac{(i+0.5)\pi}{8}x\right] \\ cos \left[ \frac{(j+0.5)\pi}{8}y\right]\bigl\}\nonumber
\end{align}
where $i=0,1,...7$, $j=0,1,...7$ and \\
\begin{equation}
  C(x) =
    \begin{cases}
      \frac{1}{2\sqrt{2}} & \text{if $x=0$}\\
      \frac{1}{2} & \text{if $x\not =0$}\\
     \end{cases}       
\end{equation}

\begin{figure*}[htp]
    \centering
    \includegraphics[scale=0.7]{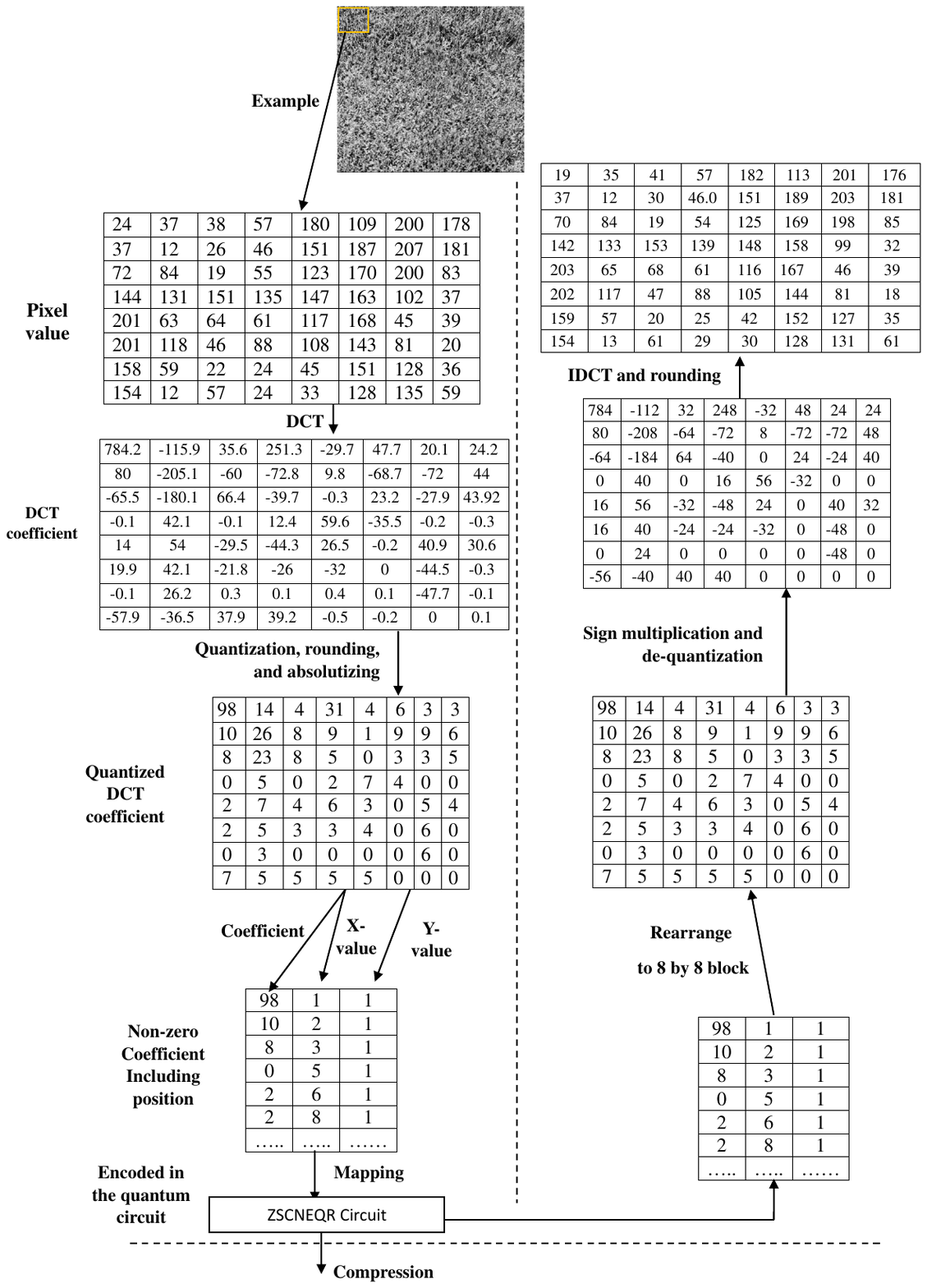}
    \caption{First block of grass image as an example of the proposed approach}
    \label{fig_exam_grass}
\end{figure*}

Figure~\ref{fig_exam_grass} exhibits the compression and decompression of the proposed approach of the first block of the 'grass' image using an 8-quantization factor which required $64\times64$ $XY$-block. It has been shown that the recovered and original blocks are not similar, which means the proposed approach is lossy compression.



\subsection{Network complexity}

In this section, the network complexity of the proposed approach is compared with the NEQR and BEC approaches to highlight the benefit of our approach. However, our approach and BEC (Boolean Expression Compression) used a pre-processing step. The complexity of the network depends on the required number of gates~\cite{li2023quantum}. For example, for position locating, one Toffoli gate (a not gate with six control gates) shown in Figure~\ref{fig_proposed_SCMNEQR_diagram} is required to locate  $X$- and $Y$-position of quantized transfer coefficient. In this work, one connection is considered a basic elementary gate, and only the complexity of the quantum part is analyzed. For generalization, the NEQR approach does not consider the time complexity measure to restore the quantum image~\cite{fan2019quantum,zhou2019quantum,chetia2021quantum}.

For a $2^n\times 2^n$ image size, and $2^{(q-1)}$ gray-scale labels, the time complexity of NEQR approach is no more than $O(qn2^{2n})$ ~\cite{bb10}. On the other hand, the time complexity of the FRQI approach is $O(2^{4n})$ for $2^n\times 2^n$ image preparation, which is too high~\cite{bb10}. BEC has been used to perform compression to minimize the time complexity of the FRQI approach. The complexity of BEC pre-processing approach is $O(2n\cdot q \cdot 2^{4n})$ ~\cite{pm}. Moreover, the compression ratio of the FRQI approach depends on image pixel value distribution. The time complexity of the BEC pre-processing approach is very high, and most of the applications are intolerable.      

The time complexity to implement $U$ quantum operation is $q+\log_2{S_X} + \log_2{S_Y}$, as seen in ~\eqref{eq6}. Here, $q$ indicates the number of non-zero transfer coefficient values after quantization, while $S_X$ and $S_Y$ represent their corresponding X and Y state connections. To implement the $\psi_2$ operation, the $U_2$ quantum operator maps the transfer coefficient values and their corresponding state connections in the quantum circuit. The entire operation in Step 2 is divided into $2^{\log_2{S_X} +\log_2{S_Y}}$ sub-operational blocks to store the transfer coefficient values of the image. When a digital image is highly disordered, the compression ratio of the FRQI approach is meager, even near zero. 

When implementing $U_{YX}$ operator, the time complexity is no more than $O(q \cdot (\log_2{S_X} + \log_2{S_Y}))$. Moreover, to control the $i^th$ qubit of the transfer coefficient value when applying $C^1_{YX}$, $\omega_{YX}^i$ uses $k$ C-NOT gates. Therefore, for every sub-operation, the time for each block is no more than $q$, which is the number of transfer coefficient values. In addition, $l$ number C-NOT is required to connect the transfer representing qubit value to its corresponding state representing qubits. The time complexity required to implement this gate is also at most $q$. In addition, to implement the reset gate operator to reset each transfer coefficient connection, the time complexity is no more than $q$ value. In summary, from the above analysis, the overall complexity of the MTGSC approach is $O[q+\log_2{S_X} +\log_2{S_Y}+q\cdot 2^{(\log_2{S_X} + \log_2{S_Y})}+q+q]$ \\ =  $O[3q+\log_2{S_X} +\log_2{S_y}+q\cdot 2^{(\log_2{S_X} +\log_2{S_Y})}]$=$O[3q+\log_2{S_X} +\log_2{S_y}+q\cdot 2^{(\log_2{S_X} +\log_2{S_Y})}]$. Therefore, the complexity of the proposed approach is no more than $O[3q+\log_2{S_X} +\log_2{S_y}+q\cdot 2^{(\log_2{S_X} +\log_2{S_Y})}]$. It has been seen that the proposed approach is less complex than the NEQR and FRQI approaches.

\section{Result and discussion}
\label{R_D}

In this section, we present the experimental results of the proposed approach, measured and analyzed using standard benchmark image datasets ~\cite{bb18,li2005object}. We compare our results to existing classical compression schemes, such as JPEG, as well as quantum domain compression methods including SCMNEQR, DCTINCQI, DCTNCQI, and DCTEFRQI. The computations were performed using MATLAB 2022b on a Windows PC with the following specifications: RAM-16GB, processor-11th Gen Intel(R) Core(TM) i7-11700@ 2.50GHz.

Additionally, we used the Quirk quantum simulator for visualization purposes. For result verification, we selected eight types of images from the database: deer, baboon, scenery, airport, building, peppers, grass, and texture ~\cite{bb17,bb18}. Details of each image can be found in Table ~\ref{image_details}.

\begin{table}
\caption{Selected image and its corresponding size}
\begin{center}
\begin{tabular}{|l|l|}
\hline
\textbf{Image} & \textbf{Image size} \\
\hline
Deer & $1024\times1024$ \\
\hline
Baboon & $512\times512$ \\
\hline
Scenery & $512\times512$ \\
\hline
Airport &  $1024\times1024$ \\
\hline
Building &  $512\times512$ \\
\hline
Peppers &  $512\times512$ \\
\hline
Grass &  $512\times512$ \\
\hline
House &  $512\times512$ \\
\hline
\end{tabular}
\label{image_details}
\end{center}
\end{table}

The compressibility of the representation approach is measured by the required number of gates per pixel map, zero and one. Zhang et al. (2013) proposed a NEQR approach that uses a frequent number of 1’s in the pixel value to map the quantum circuit ~\cite{bb10}. In the quantum circuit, 0 and 1 are implemented using a negative control gate and control gate. Besides, 0 is also considered an identity gate that does not contribute to the output. In such a way, the required number of gates per pixel to compress and represent the image maps the bit values of the transform coefficient in the quantum circuit to complete the circuit connection. In this work, the required number of gates per pixel and corresponding PSNR are considered as evaluation criteria.

\begin{figure}[!t]
\centerline{\includegraphics[width=.7\linewidth,height=6cm]{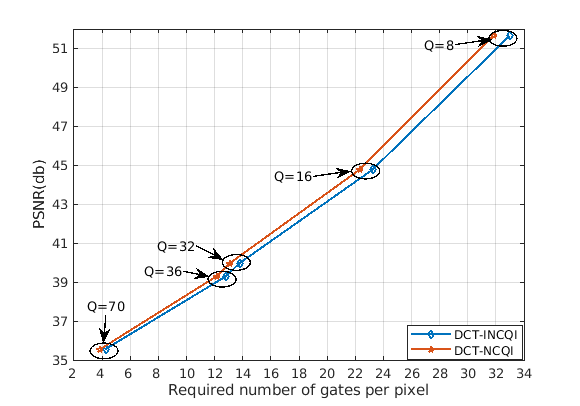}}
\caption{Computational result of the required number of gates per pixel and PSNR values of the DCTNCQI and DCTINCQI existing approaches for deer image.}
\label{RDC_DCT_INCQI_DCT_NCQI}
\end{figure}

Figure~\ref{RDC_DCT_INCQI_DCT_NCQI} illustrates the required number of gates per pixel and PSNR performance result of the DCTINCQI and DCTNCQI approaches for the deer image using 8, 16, 32, 36, and 70  quantization factors. The evaluation results indicate that the DCTINCQI approach requires a higher number of gates compared to the DCTNCQI approach. However, both methods yield the same PSNR value. It can be attributed to the fact that the DCTINCQI approach involves additional transform coefficient values, including their positions, which incur higher costs. Due to the higher number of gates required, the remaining results of these two methods are not considered.   

\begin{figure*}
    \centering
    \subfigure [deer]
    {
        \includegraphics[width=0.48\textwidth, height=6cm]{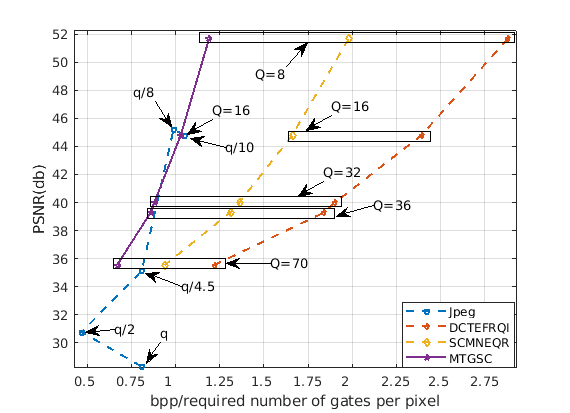}
        \label{graychannel_deer}
    }
    \subfigure[airport]
    {
        \includegraphics[width=0.48\textwidth, height=6cm ]{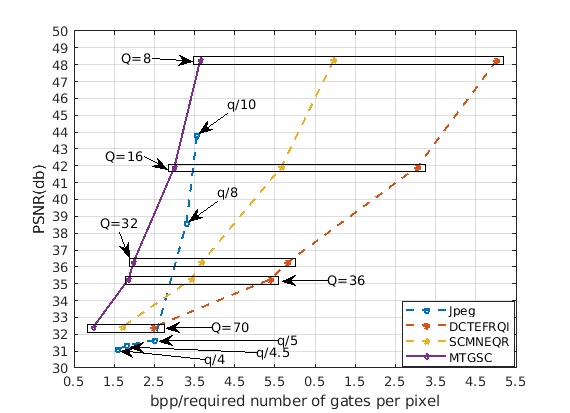}
        \label{gray_channel_airport}
    }
    \subfigure[baboon]
    {
        \includegraphics[width=0.48\textwidth, height=6cm ]{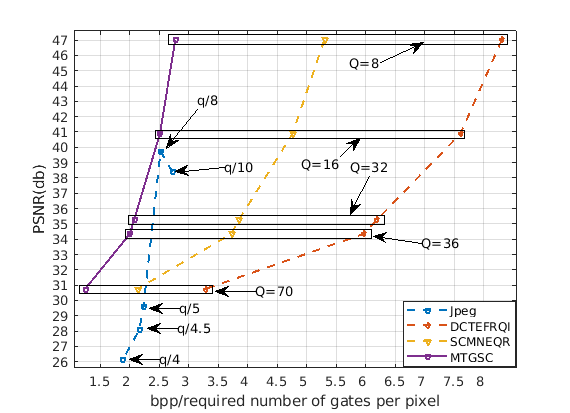}
        \label{gray__channel_baboon}
    }
    \subfigure[building]
    {
    \includegraphics[width=0.48\textwidth,height=6cm ]{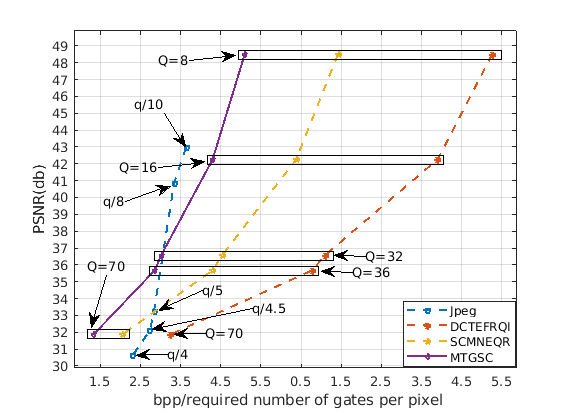}
        \label{gray__channel_building}
    }
    \subfigure[scenery]
    {
    \includegraphics[width=0.48\textwidth,height=6cm ]{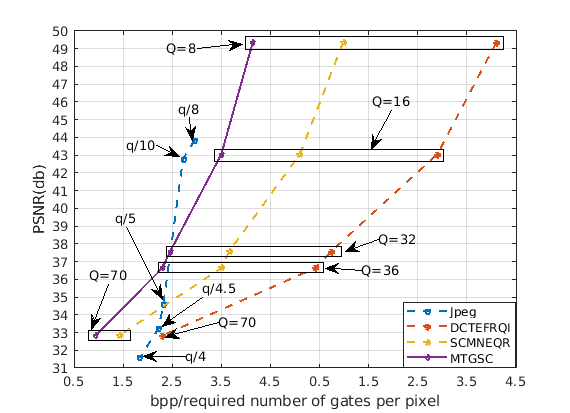}
        \label{gray_channel_scenery}
    }
    \subfigure[peppers]
    {
    \includegraphics[width=0.48\textwidth,height=6cm ]{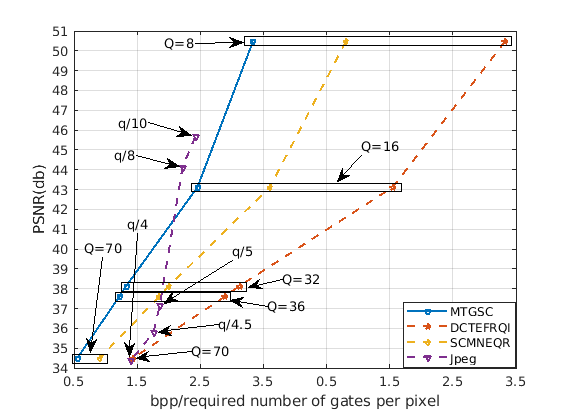}
        \label{gray_channel_peppers}
    }
    \caption{Required number of gates per pixel and PSNR metrics of the proposed (MTGSC) approach compared to Jpeg, DCTEFRQI, and SCMNEQR approaches using modified Toffoli gate and quantized transformed coefficient approach alongside novel reset gate.}
    \label{red_PSNR_comparison_color_image}
\end{figure*}

Figure~\ref{red_PSNR_comparison_color_image} shows the computed result of the required number of gates per pixel and PSNR values of the proposed approach compared to Jpeg classical compression schemes and DCTEFRQI, SCMNEQR quantum image compression approaches for grayscale deer, airport, baboon, scenery, building, and peppers images. Figure~\ref{graychannel_deer} illustrates the required number of gates per pixel versus PSNR for the deer image, comparing it with existing approaches. The comparison results show that it performs better than the Jpeg classical approach, as expected due to differences in the coding system for each quantization factor. Not being concerned with position encoding makes Jpeg unsuitable for quantum compression, as the position is critical for locating each transform coefficient value. However, compared to SCMNEQR and DCTEFRQI approaches, each quantization factor, the proposed approach outperforms in terms of the required number of gates per pixel. Based on this analysis, it is concluded that the proposed approach performs superior to DCTEFRQI and SCMNEQR quantum approaches. 

Figure~\ref{gray_channel_airport} shows the required number of gates per pixel and Peak Signal-to-Noise Ratio (PSNR) of the proposed approach applied to gray-scale airport images with quantization factors of 8, 16, 32, 36, and 70. The comparison results indicate that the proposed method performs better compression than the Jpeg approach. However, the DCTEFRQI method requires a greater number of gates per pixel than the SCMNEQR approach while achieving equivalent PSNR values. Additionally, the proposed approach outperforms both the SCMNEQR and DCTEFRQI methods in terms of the required number of gates per pixel for each quantization factor, attributed to its use of a modified Toffoli gate connection and quantized transform coefficient approach alongside a novel reset gate.

Figure~\ref{gray__channel_baboon} illustrates the relationship between the required number of gates per pixel and the PSNR values of the proposed approach for the baboon image, using 8, 16, 32, 36, and 70 quantization factors in comparison to SCMNEQR, DCTEFRQI, and Jpeg methods. The results show that the proposed approach yields values of better compression value than the Jpeg method. Conversely, the SCMNEQR and DCTEFRQI methods perform less compression means require higher quantum resources in comparison to the proposed approach across each quantization factor. These findings indicate that the proposed method compresses more efficiently than both the DCTEFRQI and SCMNEQR approaches.

Figure~\ref{gray_channel_scenery} illustrates the required number of gates and PSNR values for grayscale scenery images using the proposed approach, compared to existing state-of-the-art methods using quantization factors of 8, 16, 32, 36, and 70.  Compared to other methods such as DCTEFRQI and SCMNEQR, it shows better performance in terms of the required number of gates per pixel as a performance metric. In contrast, the required number of gates per pixel for DCTEFRQI is significantly higher than that of the proposed approach. These results are reasonable given the different encoding techniques used by each method to represent quantum image data, including the state value.

Figure~\ref{gray_channel_peppers} illustrates the results of the proposed MTGSC approach in terms of the required number of gates per pixel and the PSNR performance metric using 8, 16, 32, 36, and 70 quantization factors compared to state-of-the-art methods for the peppers image. The results indicate that MTGSC  performs better than the Jpeg approach. However, Jpeg is not suitable for quantum image compression as it does not relate to quantum circuits. A noticeable difference has been observed between the MTGSC, DCTEFRQI, and SCMNEQR approaches. Additionally, MTGSC requires a significantly lower number of gates per pixel while achieving the same PSNR value. From this analysis, it is concluded that the proposed approach compresses quantum images more efficiently than both the DCTEFRQI and SCMNEQR methods.

\begin{figure*}
    \centering
    
    \subfigure[grass image]
    {
    \includegraphics[width=0.48\textwidth,height=6cm ]{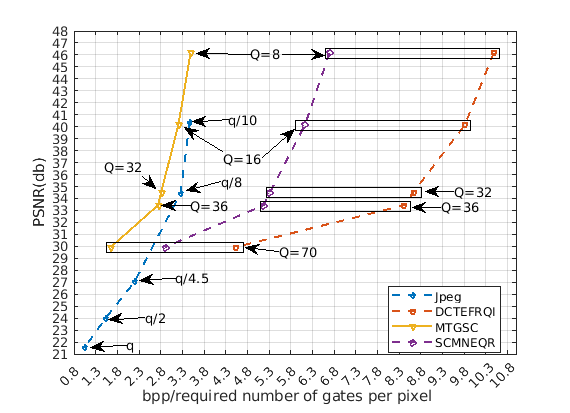}
        \label{gray_channel_texture_grass}
    }
    \subfigure[house image]
    {
    \includegraphics[width=0.48\textwidth,height=6cm ]{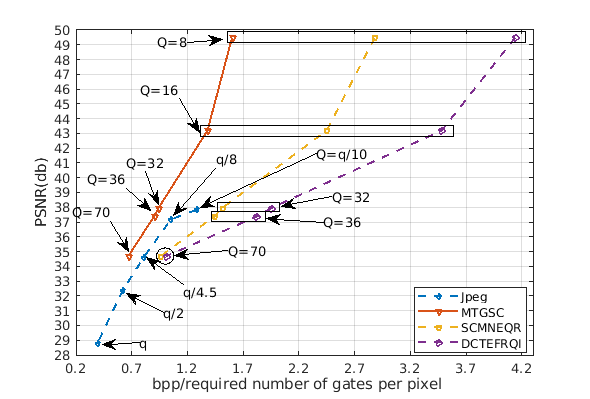}
        \label{gray_house}
    }
   \caption{Required number of gates per pixel and PSNR of the proposed approach compared to Jpeg, DCT-EFRQI, and SCMNEQR approaches for 'grass' and 'house' images using modified Toffoli gates connection and quantized transformed coefficient approach.}
   \label{gray_texture_house}
\end{figure*}

Figure~\ref{gray_channel_texture_grass} shows the proposed approach's required number of gates per pixel using different quantization factors for 'grass' images. For Q=8, it requires 3.48 gates per pixel, while SCMNEQR and DCTEFRQI draw 6.7 and 10.49. It also provides 46.18db PSNR, like DCTEFRQI and SCM-NEQR approaches. For q/10, Jpeg requires a 3.47 bpp value to transmit a 40.37db PSNR signal. For Q=16, it generates a 40.15db PSNR while it requires a 3.21 number of gates per pixel, which is less than Jpeg (3.26 bpp, q/8), DCTEFRQI (9.82), and SCMNEQR (6.13) approaches. In the case of PSNR, it generates 40.15db, which is greater than Jpeg (34.45db) but provides the same PSNR as the DCTEFRQI and SCMNEQR approach. Furthermore, for Q=32, it requires (2.81) less number of gates per pixel than SCMNEQR (5.32) and DCT-EFRQI (8.63) approaches respectively, while providing similar PSNR (34.5db). Compared to the Jpeg (2.2bpp, 27.04db) approach, it produces higher PSNR and requires more gates. In addition, Q=36 draws fewer required gates (2.74) than SCMNEQR (5.18) and DCT-EFRQI (8.41) approaches. In comparison to the Jpeg (1.52bpp, 23.95db) approach, it exhibits a higher required number of gates (2.74) and produces higher PSNR (33.43db). In addition, Q=70 requires fewer gates (1.65) compared to SCMNEQR (2.9) and DCT-EFRQI (4.53) approaches, while both draw 29.93db PSNR. Furthermore, it generates a higher value for both the required number of gates per pixel and PSNR compared to the Jpeg (1.04bpp, 21.51db) approach. In summary, comparison results show that the proposed approach performs better than Jpeg, SCNEQR, and DCTEFRQI  approaches.

Figure~\ref{gray_house} depicts the performance result of the proposed approach in case of the required number of gates per pixel and PSNR for the 'house' image using 8, 16, 32, 36, and 70 quantization factors compared to the Jpeg, SCNEQR, and DCTEFRQI approaches. For Q=8, it required a smaller number of gates (1.6) than SCMNEQR (2.88) and DCTEFRQI (4.14) approaches but generated the same PSNR (49.48) value. On the other hand, Jpeg requires less bpp (1.28) and produces less PSNR (37.83) than the MTGSC approach. For Q=16, the proposed requires 1.38 gates per pixel, which is less than the SCMNEQR (2.05) and DCTEFRQI (3.48) approaches. It provides the same PSNR (43.19db) as SCMNEQR and DCTEFRQI approaches. Compared to the Jpeg (1.05bpp, 37.17db) approach, it requires higher gates and generates a higher PSNR(43.19db). It produces a 6.02db higher PSNR while it requires 0.33 additional gates compared to the Jpeg approach. For Q=32, it required a smaller number of gates (0.94) than the SCMNEQR (1.5) and DCT-EFRQI (1.95) approaches. In the case of PSNR, it produces 37.94db PSNR like SCMNEQR and DCTEFRQI approaches. Furthermore, it requires higher gates (0.94) and generates a higher PSNR (37.94db) than Jpeg (0.81bpp, 34.62db). For Q=36, it requires fewer gates (0.9) than SCMNEQR (1.4) and DCT-EFRQI (1.82) approaches, while all approaches produce 37.37db PSNR. Compared to the Jpeg approach, it provides a 4.98db higher PSNR by occupying an additional 0.28 gates. Furthermore, Q=70 requires fewer gates per pixel (0.67) than the SCMNEQR (0.96) and DCT-EFRQI (0.91.01) approaches, but it draws the 34.69db PSNR. Moreover, it produces a higher PSNR (34.69db) than the Jpeg approach (28.75db) but requires higher gates (0.67) per pixel compared to the Jpeg (0.39bpp) approach. In summary, based on this result analysis, it is concluded that the proposed approach is more efficient than existing state-of-art approaches.  

\begin{figure*}
\centering
    \subfigure [Original]
    {
        \includegraphics[width=0.2\textwidth, height=2.5cm]{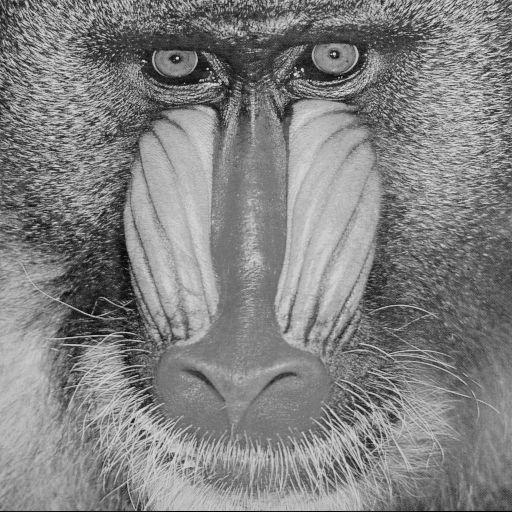}
        \label{baboon_or}
    }
    \subfigure[Rec.,Jpeg, q/4.5]
    {
      \includegraphics[width=0.2\textwidth,height=2.5cm ]{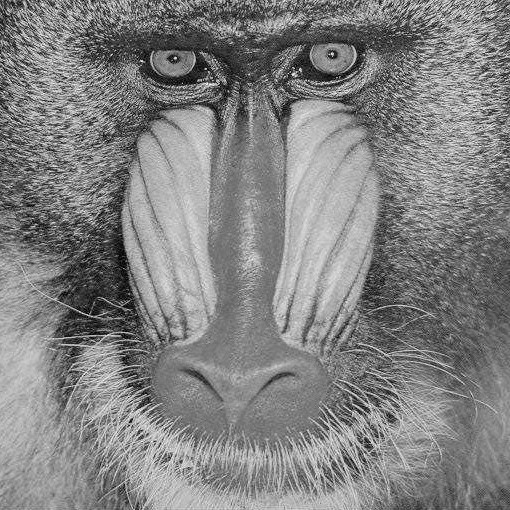}
        \label{baboon_8_jpeg}
    }
    \subfigure[Rec.,SCMNEQR,8]
    {
    \includegraphics[width=0.2\textwidth,height=2.5cm ]{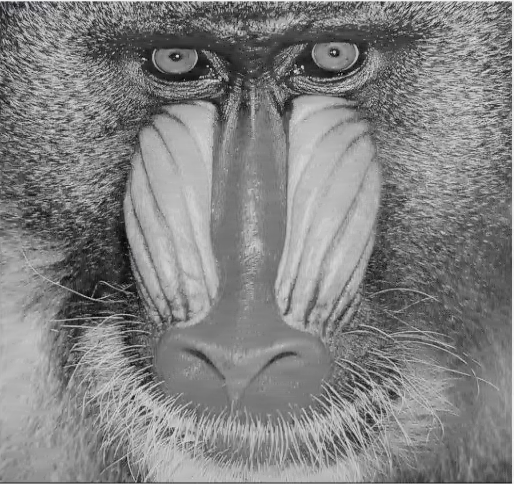}
        \label{baboon_8_baboon}
    }
        \subfigure[Rec.,MTGSC, 8]
    {
    \includegraphics[width=0.2\textwidth,height=2.5cm ]{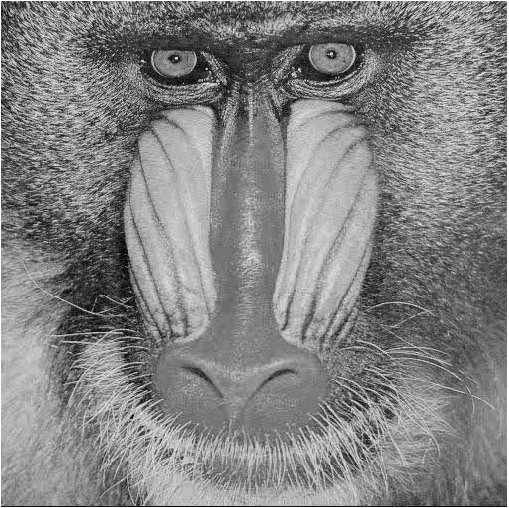}
        \label{baboon_8_ZSCNEQR}
    }
    \subfigure [Original]
    {
        \includegraphics[width=0.2\textwidth, height=2.5cm]{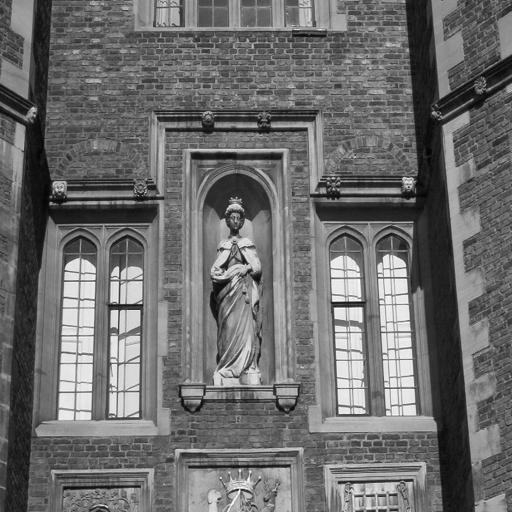}
        \label{building_gray}
    }    
    \subfigure[Rec.,Jpeg,q/5]
    {
      \includegraphics[width=0.2\textwidth,height=2.5cm ]{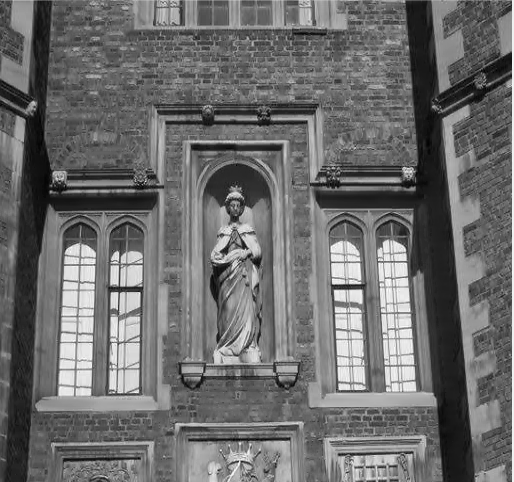}
        \label{building_jpeg_5}
    }
    \subfigure[Rec.,SCMNEQR,16]
    {
      \includegraphics[width=0.2\textwidth,height=2.5cm ]{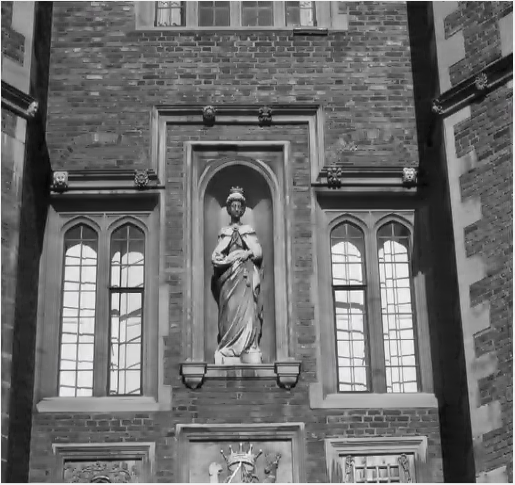}
        \label{building_32}
    }
    \subfigure[Rec.,MTGSC,16]
    {
    \includegraphics[width=0.2\textwidth,height=2.5cm ]{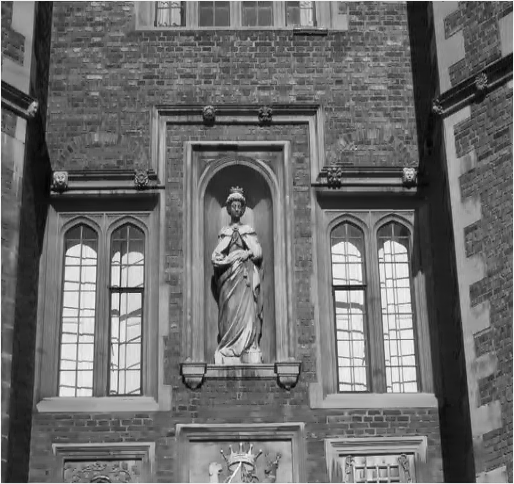}
        \label{building_16_SCNEQR}
    }
    \subfigure [Original]
    {
        \includegraphics[width=0.2\textwidth, height=2.5cm]{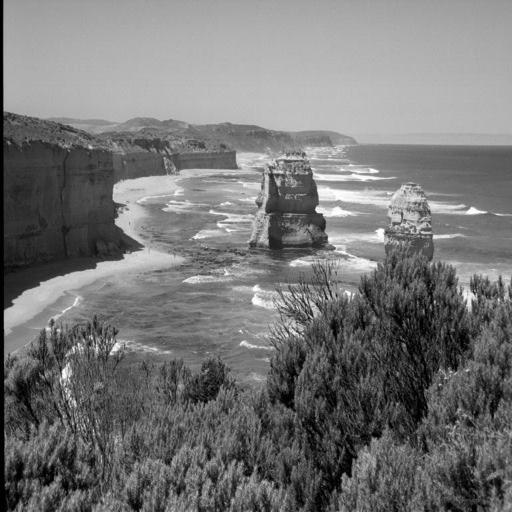}
        \label{scenery_org}
    }
    \subfigure[Rec.,Jpeg,q/1.25]
    {
      \includegraphics[width=0.2\textwidth,height=2.5cm ]{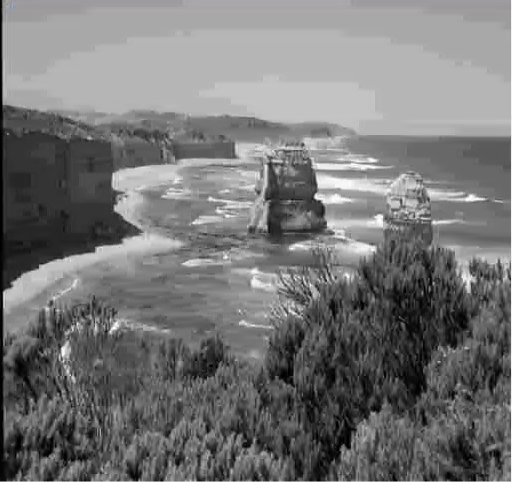}
        \label{scenery_Jpeg_1.25}
    }
        \subfigure[Rec., SCMNEQR,32]
    {
      \includegraphics[width=0.2\textwidth,height=2.5cm ]{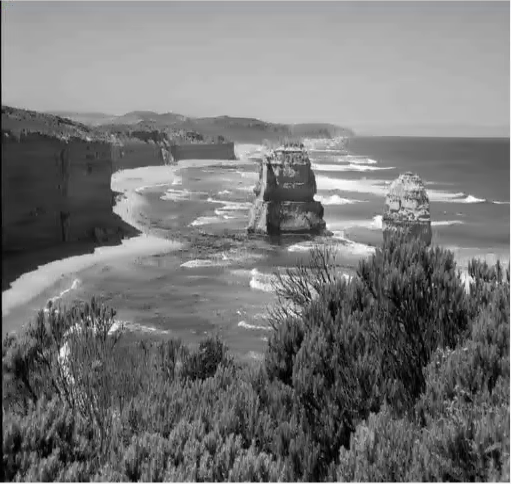}
        \label{scenery_SCMNEQR_32}
    }
    \subfigure[Rec.,MTGSC,32]
    {
    \includegraphics[width=0.2\textwidth,height=2.5cm ]{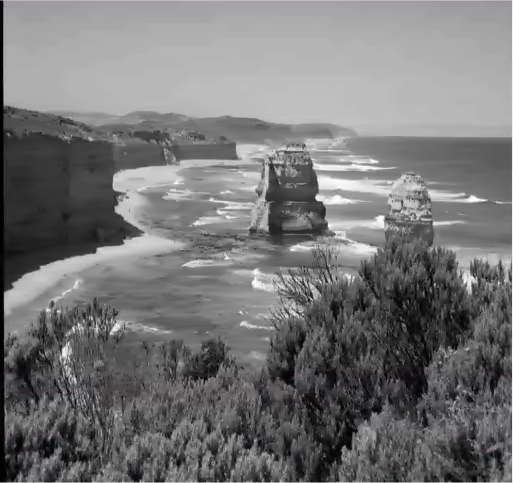}
        \label{scenery_32_ZSCNEQR}
    }
    \subfigure [Original]
    {
    \includegraphics[width=0.2\textwidth, height=2.5cm]{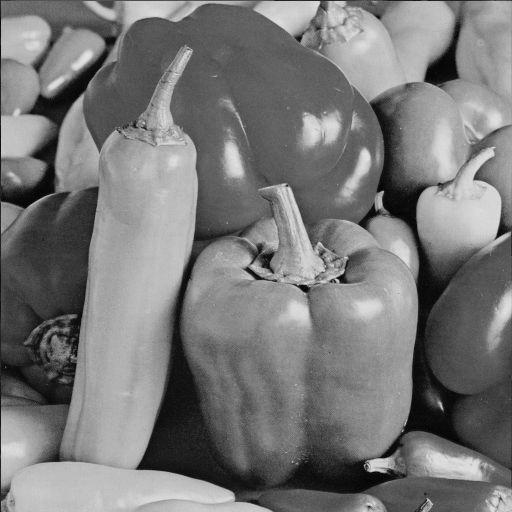}
        \label{peper_ori}
    }    
    \subfigure[Rec.,Jpeg, q]
    {
      \includegraphics[width=0.2\textwidth,height=2.5cm ]{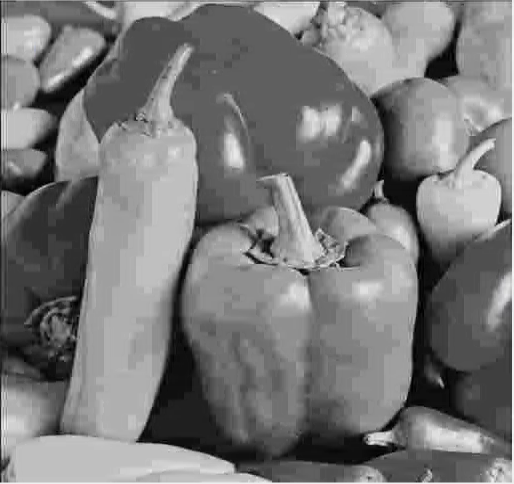}
        \label{peppers_8}
    }
    \subfigure[Rec.,SCMNEQR,70]
    {
    \includegraphics[width=0.2\textwidth,height=2.5cm ]{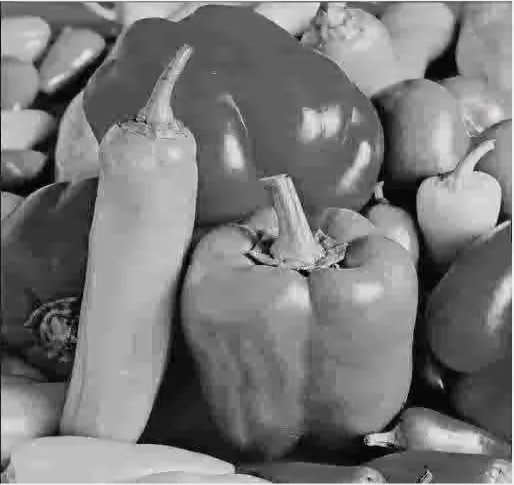}
        \label{peppers_70_scmneqqr}
    }
    \subfigure[Rec.,MTGSC, 70]
    {
    \includegraphics[width=0.2\textwidth,height=2.5cm ]{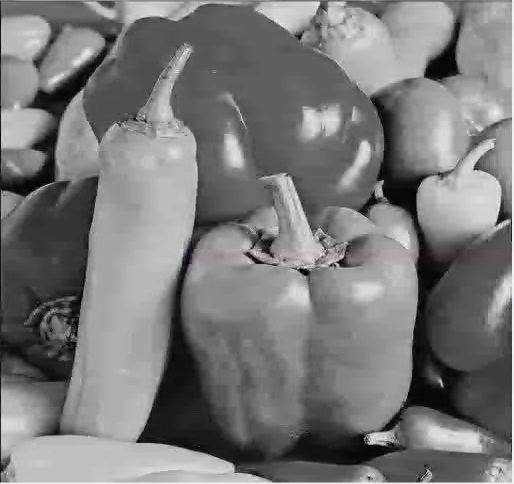}
        \label{peppers_70}
    }
        \subfigure [Original]
    {
    \includegraphics[width=0.2\textwidth, height=2.5cm]{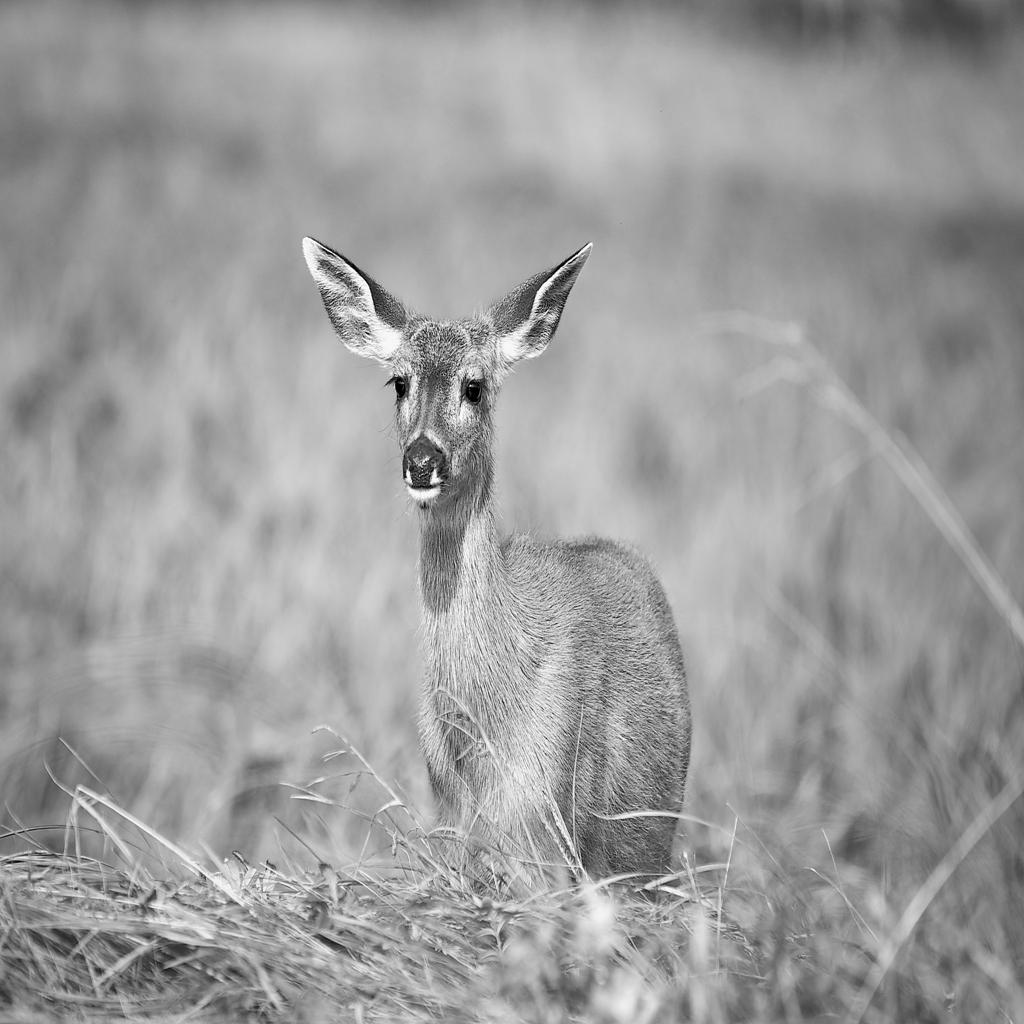}
        \label{deer_ori}
    }    
    \subfigure[Rec., Jpeg, q/2]
    {
      \includegraphics[width=0.2\textwidth,height=2.5cm ]{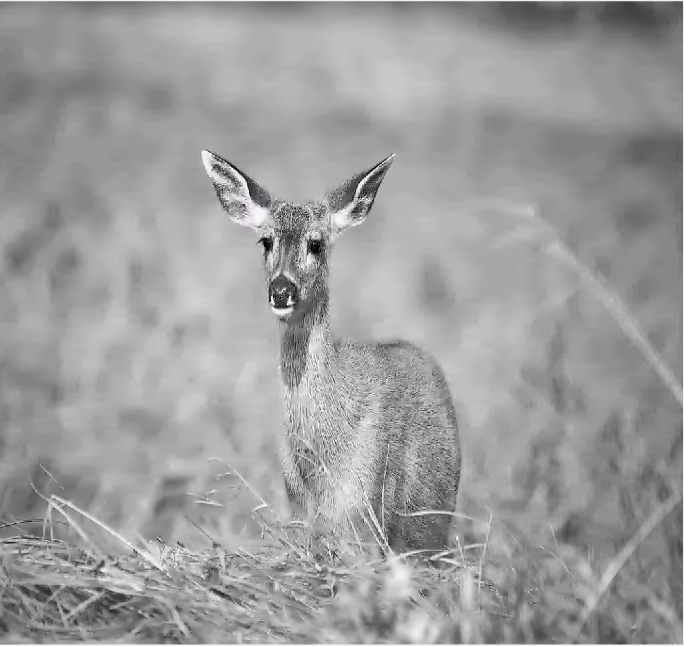}
        \label{deer_jpeg_0.5}
    }
    \subfigure[Rec.,SCMNEQR,32]
    {
    \includegraphics[width=0.2\textwidth,height=2.5cm ]{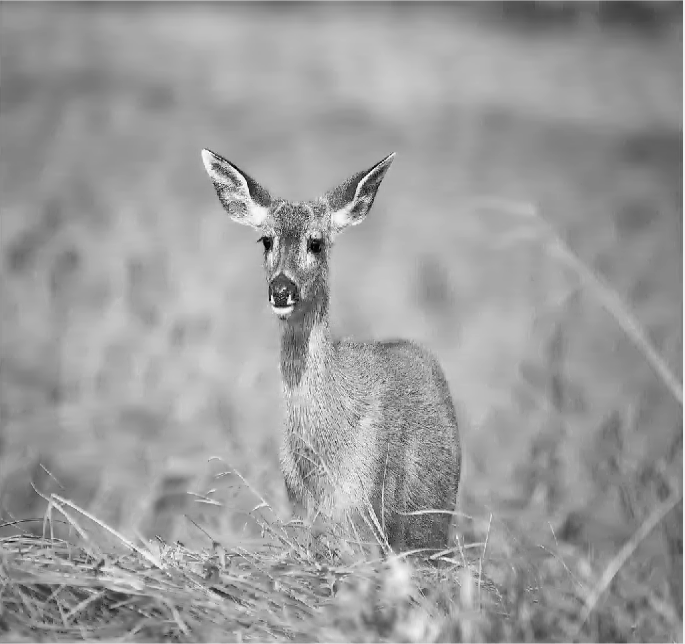}
        \label{deer_scmneqr_32}
    }
    \subfigure[Rec.,MTGSC,32]
    {
    \includegraphics[width=0.2\textwidth,height=2.5cm ]{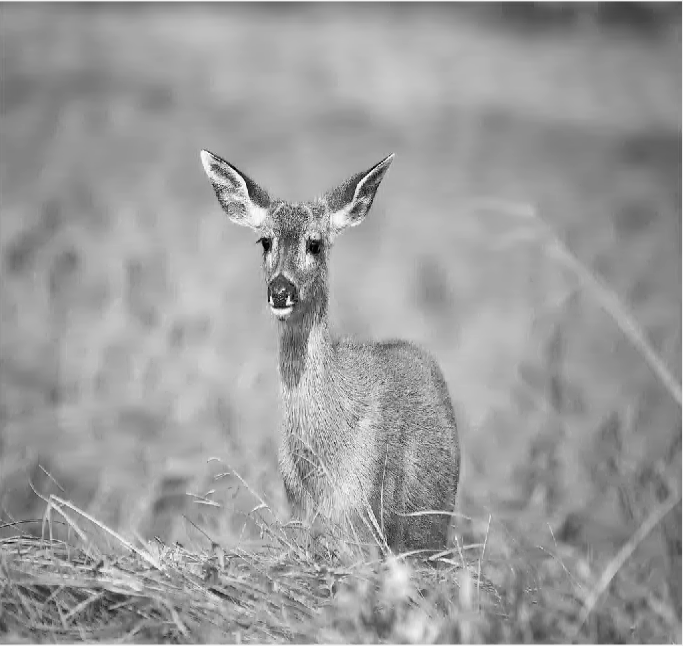}
        \label{deer_zscneqr_32}
    }
    \subfigure [Original]
    {
    \includegraphics[width=0.2\textwidth, height=2.5cm]{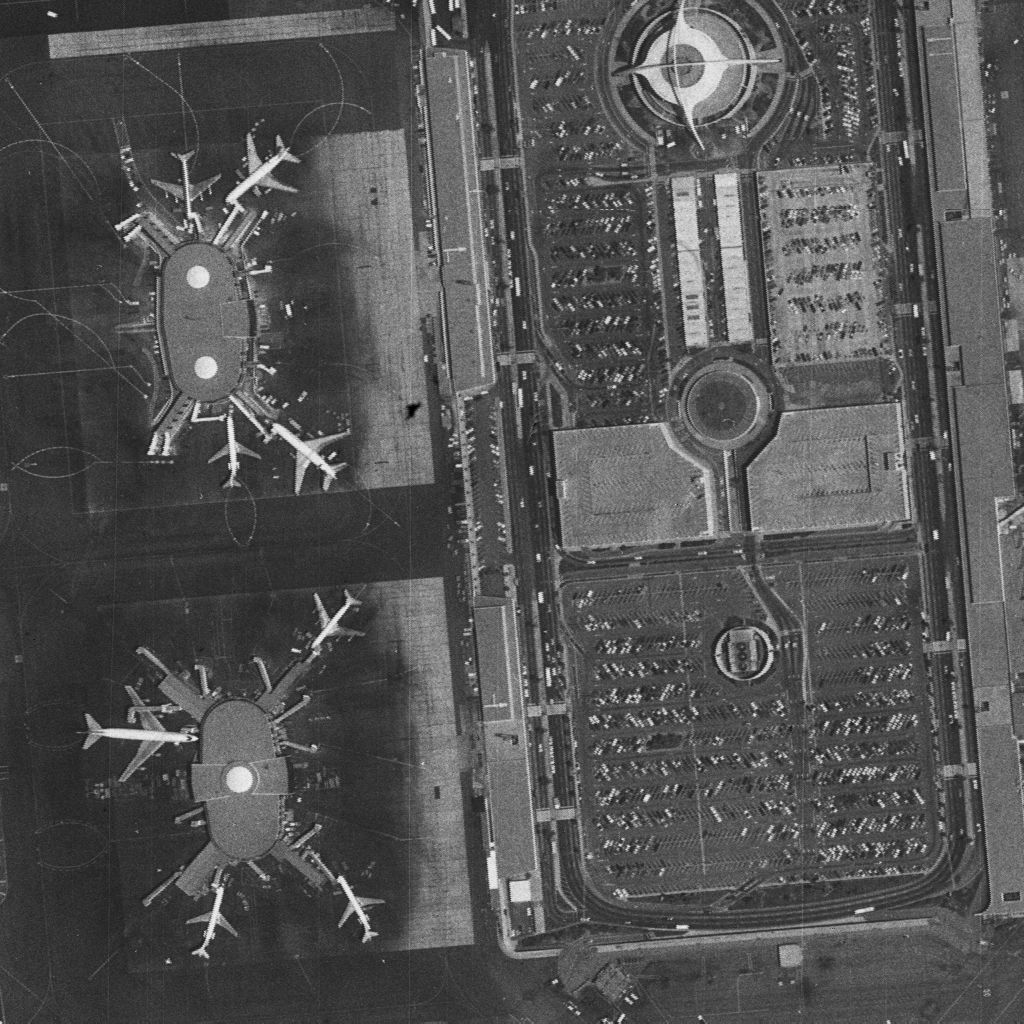}
        \label{airport_ori}
    }    
    \subfigure[Rec., q, Jpeg]
    {
      \includegraphics[width=0.2\textwidth,height=2.5cm ]{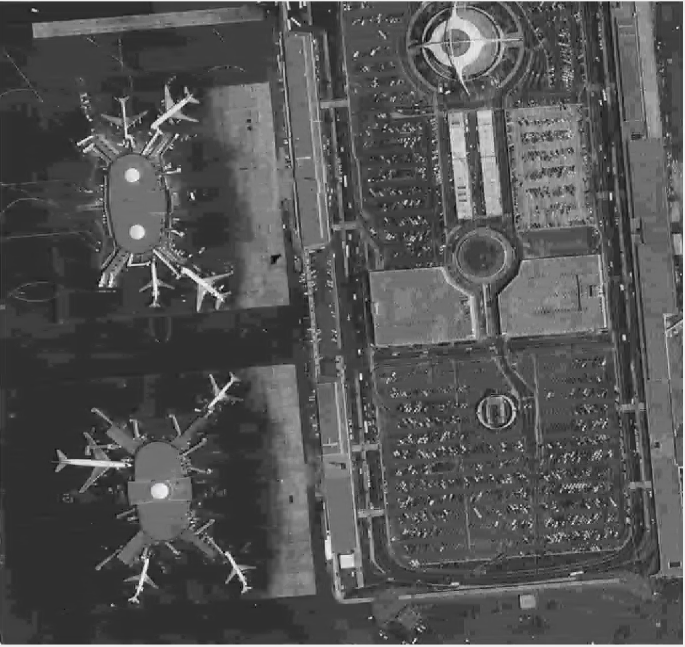}
        \label{airport_q_jpeg}
    }
    \subfigure[Rec.,SCMNEQR,32]
    {
    \includegraphics[width=0.2\textwidth,height=2.5cm ]{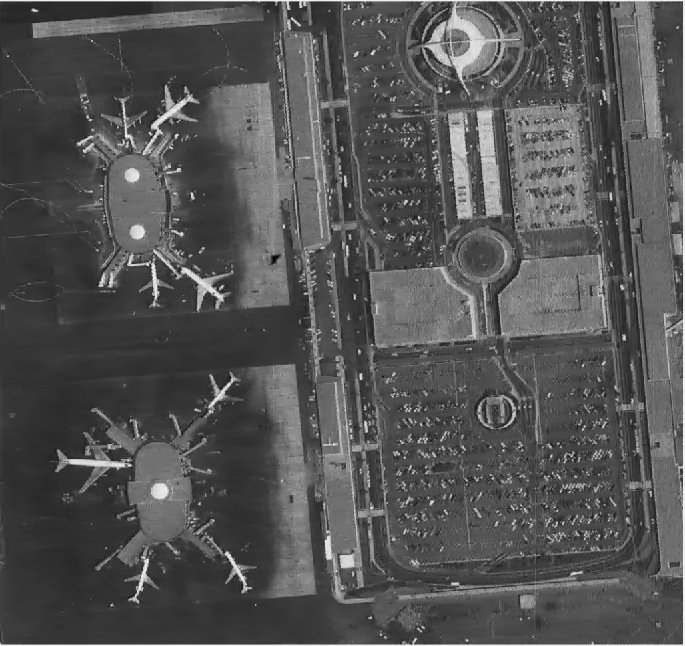}
        \label{airport_SCMNEQR_32}
    }
    \subfigure[Rec.,MTGSC,32]
    {
    \includegraphics[width=0.2\textwidth,height=2.5cm ]{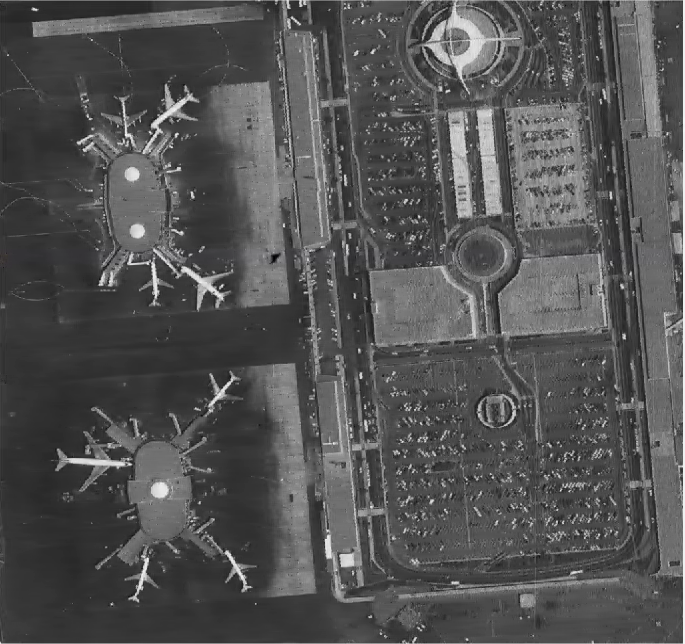}
    \label{airport_ZSCNEQR_32}
    }
    \caption{Subjective evaluation of the proposed approach for $8\times8$ image using modified Toffoli gate connection and quantized transformed coefficient approach compared to Jpeg, and SCMNEQR approaches.}
\label{subjective_evaluation_result}
\end{figure*}

\begin{figure*}
\centering
    \subfigure [Original]
    {
    \includegraphics[width=0.2\textwidth, height=3cm]{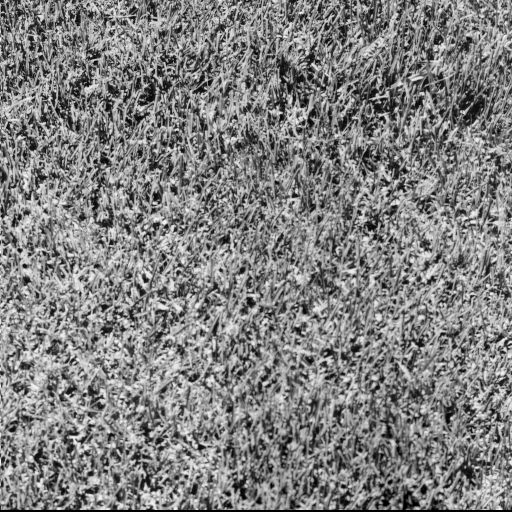}
        \label{Grass_ori}
    }    
    \subfigure[Rec.,Jpeg,q/2]
    {
      \includegraphics[width=0.2\textwidth,height=3cm ]{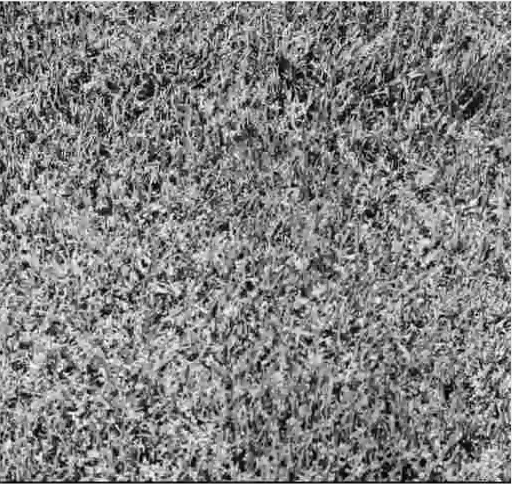}
        \label{grass_jpeg_0.5}
    }
    \subfigure[Rec.,SCMNEQR,32]
    {
    \includegraphics[width=0.2\textwidth,height=3cm ]{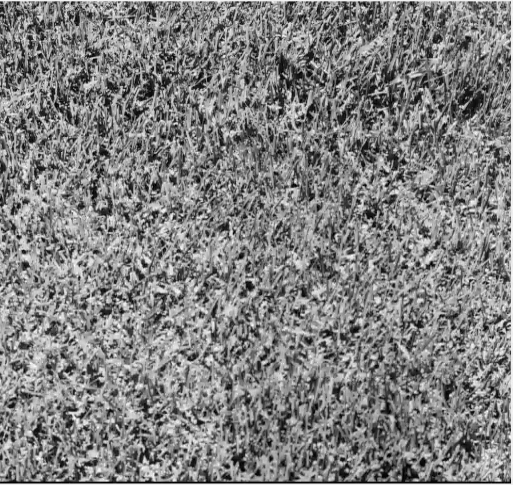}
        \label{grass_scmneqr_32}
    }
    \subfigure[Rec.,MTGSC,32]
    {
    \includegraphics[width=40mm,height=3cm]{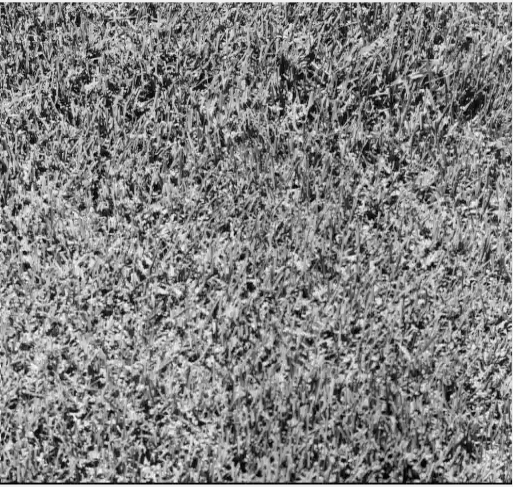}
        \label{grass_zscneqr_32}
    }
    \subfigure [Original]
    {
    \includegraphics[width=0.2\textwidth, height=3cm]{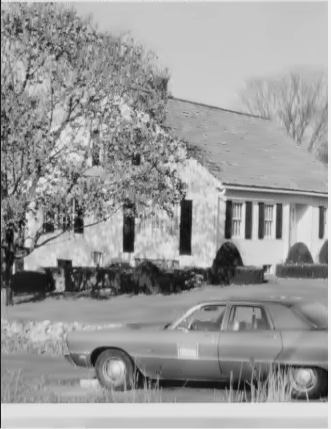}
        \label{house_ori}
    }    
    \subfigure[Rec.,q,Jpeg]
    {
      \includegraphics[width=0.2\textwidth,height=3cm ]{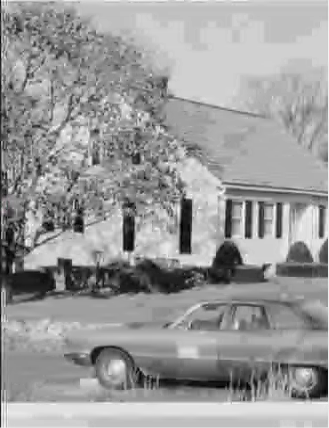}
        \label{house_q_jpeg}
    }
    \subfigure[Rec.,SCMNEQR,32]
    {
    \includegraphics[width=0.2\textwidth,height=3cm ]{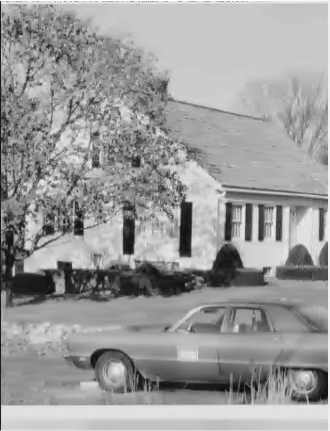}
        \label{house_SCMNEQR_32}
    }
    \subfigure[Rec.,MTGSC,32]
    {
    \includegraphics[width=40mm,height=3cm]{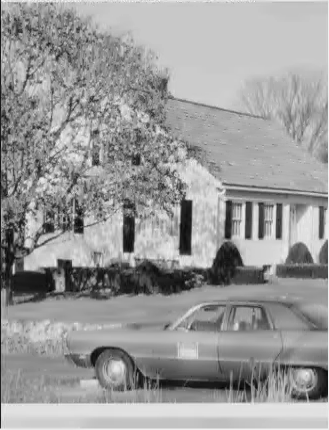}
    \label{house_ZSCNEQR_32}
    }
    \caption{Subjective evaluation of the proposed approach for 'grass' and 'house' image using modified Toffoli gates state connection approach and quantized transform coefficient approach compared to Jpeg, and SCMNEQR approaches.}
\label{subjective_evaluation_result1}
\end{figure*}

Figure~\ref{subjective_evaluation_result} shows the subjective evaluation of the original and reconstructed image that demonstrates the efficiency of the proposed approach compared to the Jpeg and SCMNEQR approaches utilizing various quantization factors and DCT pre-processing approaches. For the sake of simplicity, the results are exhibited for 8, 16, 32, and 70  quantization factors for Baboon, Building, Scenery, Peppers, Deer, and Airport images. Moreover, Figure~\ref{subjective_evaluation_result1} also demonstrates the subjective evaluation of the proposed approach for 'grass' and 'house' images using various quantization factors and DCT preprocessing approach compared to Jpeg and SCMNEQR approaches.


\section{Opportunities and Challenges}

Quantum computing improves the computational efficiency due to the properties of quantum mechanics. The pre-processing step of quantum computation is complex since it needs to create the superposition, entanglement, and parallelism properties of quantum mechanics~\cite{ruan2021quantum}. Generally, at the end of the quantum algorithm, the superposition and entanglement degenerated using the probabilistic-based amplitude of the qubit connection. Image representation and compression are becoming major fields in quantum computing. The existing algorithm involves the FRQI and NEQR quantum image representation. For digital image processing, the FRQI and NEQR approach has a lot of limitations, such as higher complexity of preparation. Moreover, how and in what way the color and its corresponding position are encoded in the quantum circuit is still a big challenge for gray-scale images. Since it stores the probability of the gray-scale information using a single qubit, it is challenging to retrieve the original information.

A quantum state is a set of a wide range of vectors performed by metrics operation in high dimensional space. Furthermore, machine learning protocols operate using the matrix operation. Adopting machine learning tasks with quantum circuits would be incredibly beneficial to simulate the natural behavior of the word. In quantum computing, complexity happens with the state preparation of pixels. In this work, we have introduced a quantized transfer coefficient block partitioning circuit and modified state connection approach to compress and represent the gray-scale image. Besides, measuring the performance metrics to demonstrate its image compression capacity using the required number of gates per pixel is introduced first. Rather than encoding pixels directly into the quantum circuit, a quantized transform version of pixels is encoded.

Based on the above discussion, a quantum circuit is the right candidate to address the higher requirement of computational resources for storing, processing, and retrieving image data for representation and compression purposes. By processing and mapping classical image data into quantum circuits, superposition, entanglement, and parallelism of a quantum state can be the advantages of the quantum properties~\cite{giovannetti2008architectures,giovannetti2008quantum}. QRAM (Quantum Random Access Memory) prepares the superposition state. It distributes the \textit{$N$$d$}-dimensional vectors into \textit{$Nd$} superposition qubit. The time complexity of the QRAM approach is $O(log(Nd))$. It needs more physical resources such as $O(Nd)$. For the required number of qubits, it increases exponentially. As a result of experimental verification, it is a question to provide the advantages or not ~\cite{aaronson2015read,adcock2015advances}. 

Noise reduction is another challenging factor for quantum image computation. In the 1990s, quantum error correction codes ~\cite{rebentrost2014quantum} were studied to improve the fault-tolerant quantum computing system ~\cite{steane1999efficient}. Nowadays, noise elimination is one of the fascinating research fields where topology quantum computing is the most concerning issue ~\cite{aasen2016milestones}. Disjunctive Normal Form (DNF) ~\cite{bshouty1995learning} is an exciting function that can learn about noisy environments where classical can not. Researchers face challenges integrating even and odd functions with DNF ~\cite{grilo2019learning}. Moreover, linearity is the nature of quantum computing. Further investigation is required to induce the non-linearity properties of quantum computing, which is another big challenge.    

\section{Conclusion}
\label{CC}

In this work, the factors responsible for increasing the complexity of state connections have been figured out through the literature survey analysis. Then, a novel MTGSC approach is proposed for various sizes of images to improve the existing SCMEFRQI, and the EFRQI approaches, making the state connection circuits more efficient. It modifies the state connection to compress the grayscale information from all transfer coefficients, including position. Moreover, the required number of gate values needs to be improved compared to the existing DCTEFRQI and SCMNEQR approaches, which means it represents and compresses gray-scale images efficiently. Another advantage is that it can reconstruct quantization images accurately due to a deterministic approach and removes the barrier of the existing approach's measurement limitations. From a network complexity analysis point of view, the proposed approach has reduced network complexity compared to the existing state of the art. The comparison results show a highly efficient compression and representation scheme for gray-scale images regarding the required number of gates per pixel and PSNR performance metric.

\section*{Acknowledgments}
The authors sincerely appreciate Charles Sturt University (CSU) for providing the RTP Scholarship
\bibliographystyle{unsrt}



\end{document}